\documentclass[11pt,a4paper]{article}
\usepackage[T2A]{fontenc}
\usepackage{amsmath,amssymb}
\usepackage{epsfig} %

\begin{document}

\bigskip

\begin{center}
\textbf{DENSITY-WAVE SPIRAL THEORIES IN THE 1960s. II.}
\end{center}

\bigskip

\begin{center}
\textbf{I. I. Pasha}\footnote{ \textit{ ${} $ email}: misterwye@mtu-net.ru} \par
\bigskip
{\footnotesize \textit{ Sternberg State Astronomical Institute, Moscow} \par
\textit{ Vavilov Institute for History of Science and Technology, Moscow }}

\bigskip
\bigskip
\bigskip

\tableofcontents

\end{center}

\newpage

\section*{Introduction\protect\footnotemark{}}
\addcontentsline{toc}{section}{Introduction}
\footnotetext{ Throughout the paper, the \textit{italicized} names in parentheses
refer to private communications as identified in the note to the list of
references.}

\bigskip

As it has been shown in the first publication under this title (Pasha 2002,
hereinafter Paper I), by the 1960s understanding the spiral structure of
galaxies entered a new stage of unusually vigorous activity, not always very
united or monothematic, but broadly grouped under the umbrella marked
``density-wave theory''. Its foremost enthusiast and proponent was C.C. Lin.
His papers with Frank Shu (Lin and Shu 1964, 1966) had a big and immediate
impact upon astronomers, at least as a welcome sign that genuine
understanding of the spiral phenomenon seemed in some sense to be just
around the corner. Already at the time, however, Lin's optimism for spirals
as \textit{quasi-steady} waves was not entirely shared by other experts, and toward the 1960s it
had become very clear to everyone that much hard work still remained to
explain even the persistence, much less the dynamical origins, of the
variety of spirals that we observe.

\bigskip

We start this second part of our narrative with the events that occurred and
developed right in the period of Lin and Shu's initial semi-empirical
explorations of 1963-66 on the alternative, dynamical front of
\textit{sheared}-wave research. It was those early analyses that first taught us, then in a
local approximation, that massive shearing disks tend to be wonderful
amplifiers and to respond strongly, though always in a trailing-spiral
manner, to several quite plausible forms of forcing. After that, we will use
Chapter II to describe most engaging topics like neutral tightly wound
modes, spiral shocks and star migration that Lin and Shu plus several
associates continued to explore from about 1966 onward, whereas in Chapter
III we will turn to a fascinating and very serious difficulty with the group
velocity that emerged only near the end of that decade. We will try to wrap
it all up in Chapter IV which will focus mainly on a remarkable conference
on spiral structure that took place in Basel, Switzerland in August 1969.
Though its coverage may have been a little too slanted a priori toward
praising mostly just the Lin-Shu ideas as major steps forward in this
subject, that meeting also attracted nearly all of the other main players,
and it appears interesting now to examine in retrospect which points they
themselves chose to emphasize there.$^{} $

\newpage

\section*{I. ORIGINS OF SWING AMPLIFICATION}
\addcontentsline{toc}{section}{I. Origins of swing amplification}

\bigskip

{\footnotesize \begin{list}{}{\leftmargin4cm}
\item \textit{W. Heisenberg}: How certain is it that the spirals are permanent structures? May it not
rather be a process of continuous formation? Spiral structure might be very
quickly washed out by rotation, but new spirals could be formed by
fluctuation of density.

\item \textit{J.H. Oort}: I agree. But it is difficult to conceive how spiral structures which
extend over an entire galaxy could be formed entirely anew at intervals of
one or two revolutions of the galaxy.
\begin{flushright}
\textit{Oort 1965, p.23}
\end{flushright}

\ldots then a spiral arm is some sort of a wave. Once one says this, of
course, one runs into an enormous number of possibilities.
\begin{flushright}
\textit{Prendergast 1967, p.304}
\end{flushright}
\end{list}}

\bigskip

\subsection*{1.1 Cambridge union}
\addcontentsline{toc}{subsection}{1.1 \it Cambridge union}

\bigskip

{\footnotesize \begin{list}{}{\leftmargin4cm}
\item Since Lord Rosse discovered spiral structure in M51 the explanation of this
beautiful form has been one of the outstanding problems of cosmogony. The
straightforward belief that this structure is a natural consequence of a
swirling motion was probably held by many of the early observers and it is
our hope that the present work goes some distance to establish that belief
on a firm theoretical foundation.
\begin{flushright}
\textit{Goldreich {\&} Lynden-Bell 1965b, p.125}
\end{flushright}
\end{list}}

\bigskip

\noindent Donald Lynden-Bell and Peter Goldreich met in the fall of 1963 in Cambridge,
UK. One of them had been back there from the USA for a year already, as a
University lecturer in mathematics and director of mathematical studies in
Clare College,\footnote{ Once he got his PhD degree in 1960, Lynden-Bell
left Cambridge for California. Working at Caltech with Sandage, he solved
problems on isolating integrals of motion (see Lynden-Bell 1962) and also
made, with Sandage and Eggen, the classical work on high-velocity (old)
stars (Eggen et al 1962) that proved the fact old-time contraction of our
Galaxy. Besides, he visited Chandrasekhar at Yerkes Observatory for
large-scale instability problems.} the other had just arrived on Ю one-year
US National Research Council postdoctoral fellowship.\footnote{ Goldreich's
1963 thesis was on planetary dynamics.} Goldreich had a concept of actual
problems in galaxy dynamics and no plan to pursue them, but Lynden-Bell got
him captivated by the prospects of spiral regeneration.\footnote{ ``What I
knew about spiral structure I learned from a course at Cornell entitled
``Cosmology and Evolution'' that I took in the winter of 1962 while still a
graduate student. It was the only astronomy course I ever took. (After
completing my thesis, I was appointed an instructor and taught the course
the following year.) From this course I learned that young stars were
concentrated in spiral arms and became aware of the winding problem. [\ldots
] My thesis advisor Thomas Gold mentioned Donald Lynden-Bell's name to me as
someone who did interesting work on stellar dynamics. Otherwise I didn't
know anything about him before arriving in Cambridge. Nor did I have any
intention of working with him. I cannot recall how and why we started to
collaborate, but probably it was due in large part to Donald's infectious
enthusiasm for pretty much any topic in astronomy or related fields''.
(\textit{Goldreich}) \par ``I think my enthusiasm was that the stability of a differentially
rotating disk even one modeled as gas had not been worked out and understood
and our mathematics should allow us to understand that problem''.
(\textit{Lynden-Bell})} Enthusiasm and youth -- Lynden-Bell was 28, Goldreich was 24 -- tempted
the Cambridge researchers by the confidence that this trail would lead to
the solution of the old great puzzle, and they started marching on the
``spiral arms as sheared gravitational instabilities'' (Goldreich {\&}
Lynden-Bell 1965b, hereinafter GLB) with a salvo of ``requirements of any
theory''.

\begin{quotation} {\footnotesize
\noindent ``Any theory must be wide enough to contain the bewildering variety of
galactic forms. The conventional picture of two spiral arms starting
symmetrically from the nucleus and winding several times around like
continuous threads is wrong in several aspects. In only about a third of all
normal spirals can it be claimed that just two arms are dominant and
although in these there is some tendency to symmetry it is not always very
pronounced. [\ldots ] The remaining two-thirds of normal galaxies are
multiple armed structures. In Sc's the arms often branch at unlikely angles
and the whole structure is considerably more messy than the conventional
picture. A swirling hotch-potch of pieces of spiral arms is a reasonably apt
description. A correct theory must have room for neat symmetrical two-armed
spirals, but it must not predict that most normal galaxies should be like
that. The mechanism of spiral arm formation must be so universal that it can
still work under the difficult messy conditions of a typical spiral galaxy''
(GLB, p.126). \par}
\end{quotation}

No less categorical was the authors' view of the acute `winding problem'
raised just a few years ago (Prendergast {\&} Burbidge 1960; Oort 1962) to
strengthen the evidence that ``anything in the Galaxy is sheared at such a
rate that at the end of perhaps one or two rotation periods it will be quite
unrecognizable'' (Prendergast 1967, p.304).

\begin{quotation} {\footnotesize
\noindent ``Unless the galaxies have conspired all to be spiral together for a very
brief period we must deduce that either (1) the spiral structure rotates
nearly uniformly although the material rotates differentially, or (2) the
arms are short-lived but reform as open structures, or (3) that the
observations are wrong and spirals rotate nearly uniformly'' (GLB, p.127). \par}
\end{quotation}

``To admit (3), is to say that the theorist is bankrupt of ideas'', -- GLB
judged (p.127); definitely higher they favored ``perhaps the most promising
of the theories based on (1)'' that was being made across the Atlantic (Lin
and Shu 1964),\footnote{ It is not entirely clear when and how GLB had
first learned about Lin's spiral interests and initial steps. Lynden-Bell
does not think they had ``any thoughts about Lin or about steady waves''
when they worked in 1963 (\textit{Lynden-Bell}). But soon afterwards they knew about the Lin
{\&} Shu 1964 paper from its preprint that Lin had sent to Lynden-Bell in
mid-July 1964 to acknowledge his own receipt of the GLB preprints. ``My
reaction to that paper was that Lin and Shu had missed out the real problem
by leaving out the pressure. While I read that paper my feeling was that had
I been sent it to referee I would have rejected it. [\ldots ] I believe that
if the paper of Lin and Shu had not been written we would have written
essentially the same paper, and I think [one has] the information to deduce
that from [\ldots ] my thesis along with our GLB paper, and one is a natural
outcome from the other and the more detailed stability calculation.''
(\textit{Lynden-Bell})}  yet what they found even more consistent with the sheer complexity of
actual galaxies was their own ``second type of theory''.

\bigskip

Spiral arms, the authors reasoned, are recognized above all by their
brightness due to hot massive stars that are being formed there. For all
that formative period, considerable compression of interstellar gas is
needed. It logically calls for Jeans instability as occurring mostly in the
spiral arms, and ``this at once raises the question whether the arms
themselves can be due to gravitational instability on a slightly grander
scale'' (GLB, p.126).

\bigskip
\bigskip

\subsection*{1.2 Swing amplification}
\addcontentsline{toc}{subsection}{1.2 \it Swing amplification}

\bigskip

{\footnotesize \begin{list}{}{\leftmargin4cm}
\item In the severe amplification, Goldreich and Lynden-Bell offered one real
nugget of a discovery.\footnote{ The now accepted term \textit{swing amplification}
had been introduced not in the original GLB and JT papers of the mid-1960s, but
some 15 years later, in one of Toomre's conference talks (Toomre 1981).}
\begin{flushright}
\textit{Toomre 1977, p.474}
\end{flushright}
\end{list}}

\bigskip

\noindent Lynden-Bell (1960) had already tried to materialize his
spiral-regeneration idea. He was then riveted to strict
modal analyses of gas sheets which applied in the case of rigid rotation
only, but he hoped their sensible modifications would nevertheless give him
a correct view of the effects of shear. He found this way misleading, yet he
retained his original interest.\footnote{ ``I was already at work on the
spiral problem in 1959-60, and an outline of the changing wavelength
stabilizing modes as they get sheared is given in my thesis \textit{with the
deduction that probably this theory of spiral structure will not work}. One chapter
was nevertheless entitled ``Towards a regenerative theory of spiral arms.''
(Lynden-Bell 1964d)} Goldreich told Lynden-Bell when they met that Gold, his
thesis advisor, had imagined some such concept, too, but could not work it
out.\footnote{ ``I imagine we are but the present end of a long line of
people who believed these ideas'', Lynden-Bell reacted (1964d).} He also
told that with Gold's influence his own reflections on the spiral-winding
problem and the fact that young stars are concentrated in the arms had set
him to thinking about \textit{local} gravitational instability in differentially rotating
disks. Lynden-Bell immediately appreciated this small-scale approach, and
they took it up together.

\bigskip

Imagine a washboard-like sinusoidal disturbance with its parallel crests and
troughs oriented initially at some arbitrary, perhaps even `leading', angle
with respect to the galactocentric direction. The basic differential
rotation of that region of a galactic disk will slide (shear, swing) that
density pattern as if it were painted material, except that the sinusoidal
disturbance is itself a wave and its own amplitude will evolve amidst the
shearing. To explore the actual character of this time evolution, GLB
figured out a neat 2$^{nd}$-order differential equation (almost as if for a
mass vibrating on a string, though now with a time-variable spring rate). In
that way that they \textit{discovered} that such special waves can get amplified very strongly
indeed as the shear sweeps them around through the fully open orientation,
especially in the case when their gaseous equivalent (Goldreich {\&}
Lynden-Bell 1965a) to Toomre's $Q$ is as low as unity and the azimuthal
wavelength $\lambda _{y} $ matches Toomre's (1964a) axisymmetric critical
$\lambda _{T} $ (Fig.1).

\begin{figure}
\centerline{\epsfxsize=0.9\textwidth\epsfbox{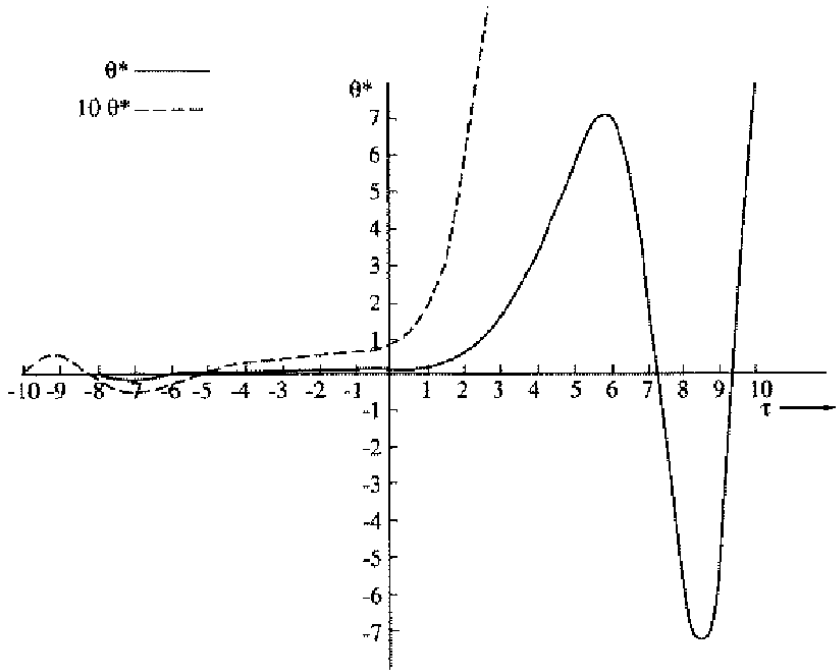}}
\caption{\footnotesize \textit {Amplitude amplification of a wave in the course of its
swinging from leading $(\tau < 0)$ to trailing $(\tau > 0)$}.
(The figure is reproduced from Goldreich
\& Lynden-Bell 1965)}
\end{figure}

\bigskip

Specifically, GLB considered a patch of a gravitating gas sheet, small and
distant from the rotation axis to allow rectangular geometry and neglect
radial variability of all its characteristics except angular speed $\Omega
(r)$ defining the shear rate $A = - \raise.5ex\hbox{$\scriptstyle
1$}\kern-.1em/ \kern-.15em\lower.25ex\hbox{$\scriptstyle 2$} $\textit{rd}$\Omega
$\textit{/dr}. They attached co-moving axes $x$ and $y$, oriented one radially and the other
along the flow, to shearing material and explored in new axes ${x}' = x,$
${y}' = y + 2Axt$ wave harmonics of the form $\exp [i(k_{x} {x}' + k_{y}
{y}')]=\exp [ik_{y} (y - \tau x)]$. Each of them knew its invariable
azimuthal wavenumber $k_{y} $ and turned by the shear `clock hand' $\tau =
2At - k_{x} / k_{y} $ (pointing radially at $\tau = 0)$ in the amplitude
control of an inhomogeneous in $\tau $ differential equation.\footnote{ ``I
remember Peter coming into my room saying he had an interest in spiral
structure and that he did not know how to solve the problem but had figured
out what coordinates to use. He then told me about his shearing coordinates
which were the key to that work.'' (\textit{Lynden-Bell})\par ``I certainly didn't solve
anything substantial, but I believe that I recognized that these coordinates
exchanged homogeneity in time for that in $x$. This was probably the most
important contribution I brought to my collaboration with Donald. [\ldots ]
I don't know whether shearing coordinates had been used in fluid problems
before GLB. However, they are such a normal choice that it would surprise me
if that had not been.'' (\textit{Goldreich})\par The sheared disturbances were adopted in the
first papers on the non-axisymmetric local dynamics of galactic disks (GLB;
Julian {\&} Toomre 1966). More classical, separable forms $A(r)\exp [i\omega
t - m\theta )]$ were recognized in the spiral-mode context just one or two
years earlier (Lin {\&} Shu 1964; Hunter 1963; Kalnajs 1963). (Lindblad had
long ago been using them in his cumbersome bar-spiral theories. Contopoulos
recalls (\textit{Contopoulos}) that when in 1962 he told Lin about Lindblad's work in some
detail, Lin took with him to MIT a batch of the latter's articles but then
confessed that he did not understand them and preferred to start working
from the scratch.) The interconnection of these two types of disturbances
was for years questionable, and, for instance, Hunter who commented on it in
his 1972 review criticized the ``seemingly arbitrary decision'' about using
shearing axes and their related Fourier-analysis, pointing out that although
it leads to solutions ``that have certain desirable properties'' it yet
``does not show why these particular solutions should be especially
significant'' and, after all, ``masks the possibility of steady waves''
(Hunter 1972, p.234-35). Lynden-Bell admits that he ``did not see how to
translate our result into a real stability result on $\exp (i\omega t)$modes
with $\omega $ real or complex'', referring to Drury (1980) as one who first
``showed how to do this'' (\textit{Lynden-Bell}). However that may have been, sheared techniques
adequately captured a very powerful amplification process, and this alone
was to sound an alert to the danger of underrating them.}  $^{} $In its
structure, shown by GLB to be the same for infinite and finite thickness
models,\footnote{ The reason why GLB neglected much simpler but sufficient
``thin disk models with infinite density'' is curious. They believed that
those were ``violently unstable since the growth rate of Jeans'
gravitational instability is proportional to $(G\rho )^{1 / 2}$'' (GLB,
p.127). Lynden-Bell disclosed the misthought once he had submitted the
paper. Yet he did not disavow it by inserting corrections in proof: that
would indicate an obvious inelegance of the authors' original analyses, and
to reduce that would have meant to redo the whole publication because quite
a number of its key discussions leaned principally on the vertical, third
dimension. \par \textit{Lynden-Bell to Toomre}: ``I have now read properly your work (Toomre 1964a) and
write to apologize for our tirades against infinitely thin disks. Earlier we
held the belief that because of the form of Jeans' instability formula all
bodies of infinite density must be unstable with infinitely rapid growth
rates and that analyses that only found finite rates were not really
treating true Jeans instability but rather the associated divergence-free or
incompressible oscillations of a compressible fluid. [\ldots ] I now agree
that sufficiently anisotropic velocities can change the look of Jeans'
criterion so that you are really discussing the same instabilities that we
are.'' (Lynden-Bell 1964b)\par ``The centerpiece equation of GLB is more
complicated than necessary because we did not clearly realize that it would
have been adequate to study two-dimensional sheets. Thin disks faithfully
capture all the horizontal dynamics of thick ones in the context of density
wave theory and swing amplification. All of the regenerative spiral
structure story could have been told in the context of two-dimensional disks
because, aside from a minor correction for vertical thickness, all that
matters for the dynamics is the horizontal velocity dispersion.''
(\textit{Goldreich})} they read the behavioral scheme of such waves. At initial stages of
their leading orientations the inter-crest spacing $\lambda $ is small and
gas pressure ensures stability. As the waves are swept round, $\lambda $
rises (right up to $\lambda _{y} )$, the pressure loses its effect, and the
net shear comes into play. It tends to feed genuinely `well-organized' gas
perturbations, and the waves get amplified.\footnote{ To better understand
the amplifying mechanism, GLB first looked into the situation with an
infinite medium subjected at some moment to a slight disturbance in the
plane of rotation, its gravity and pressure being turned off. Each
elementary `fluid volume' there starts moving along its epicyclic orbit
``with only one velocity at each point at any one time'' (unlike its stellar
counterpart with no oscillatory phase correlation of its stars), but as the
epicyclic period changes with radius due to shear, the motions of `perturbed
elements' on each given azimuth are progressively more and more out of
phase, the stronger the larger is their separation, because of which the
density amplitude in places grows with time. In the non-axisymmetric case,
azimuthal phase (and amplitude) dependence of initial perturbed motions
becomes one more growth factor: the shear brings parts with different phase
and slightly different radius to a common azimuth, which only adds to the
proportional phase-difference dependence on the radial distance and makes
the density growth still stronger. But the same shear also produces a
counteraction as it shortens trailing wavelengths thus reducing amplifying
capacities. Overall, consequently, ``this interesting behavior is not
directly related to spiral arm formation.'' (GLB, p.130)} But by the time of
their considerable trailing there comes the renewed dominance of the
pressure term and, with it, renewed oscillation, now at a largely enhanced
amplitude.\footnote{ ``I suspect we didn't expend much effort attempting to
provide a physical as opposed to a mathematical explanation for the
transient growth of sheared waves.'' (\textit{Goldreich})}

\bigskip
\bigskip

\subsection*{1.3 Spiral regeneration, take two}
\addcontentsline{toc}{subsection}{1.3 \it Spiral regeneration, take two}

\bigskip

{\footnotesize \begin{list}{}{\leftmargin4cm}
\item In order to continue the problem you must then do something nonlinear, or
you must simply publish the results. The authors mentioned did something
non-linear.
\begin{flushright}
\textit{Prendergast, 1967, p.309}
\end{flushright}
\end{list}}

\bigskip

\noindent The discovery of strong amplification of shearing formations in
self-gravi\-tating systems must have surprised most of the cosmogonists and
dynamicists who used to think of galaxies as figures of basically uniform
rotation. In essence, with this elementary and natural `microprocess'
Goldreich and Lynden-Bell struck upon a powerful engine for generating
spiral structures, at least in a transient way. Still what they had dealt
with so far was a wave-\textit{propagation} problem that gave no closed dynamical picture, even
in local setting. It left untouched the vital points of fresh-wave sources
(one-time, periodic or permanent; external or internal; distributed or
compact) and resulting responses. This engaged our authors throughout much
of the remainder of their work where they tried to build a home by way of
``reasonable speculation which [they] probably felt was justified by [their]
solid result'' (\textit{Goldreich}).

\bigskip

Within this speculation, ``the predicted return to oscillatory character
need not occur''. With isothermal gas, it gets ``energetically
advantageous'' and ``energetically possible for the nonlinear modes to
continue to condense rather than to revert to oscillatory behavior'' (GLB,
pp.139, 150), because energy released during the gravitational collapse
cannot be stored as thermal energy and is radiated away.\footnote{ One knew
well the interstellar gas as being heated up by star-formation regions and
particularly by supernovae, but also dissipative, tending to self-cooling
and forming clumps at least slightly bounded by self-gravity. This invited a
no less than two-component gas scheme with molecular clouds as its discrete,
dense, cold and inelastic part. Such a mix badly approximates to an
isothermal gas sheet, however, and it carries over no better to acoustic
waves. ``It is not clear at all how one may go about describing the
collective behavior of such a medium -- Kalnajs reasoned. -- Clearly an
application of the hydrodynamical equations (including magnetic effects),
correct in principle though, leads to a problem of unmanageable
proportions'' (Kalnajs 1965, p.56). He thus ``pretty much avoided gas
dynamics.'' (\textit{Kalnajs}) Toomre had taken some such action as if continuing his
star-disk-stability study, and even submitted a special paper to ApJ
(Toomre 1965), but then he dropped it suddenly and was never upset
about having retreated (\textit{Toomre}). In contrast, Goldreich and Lynden-Bell, who
claimed priority to gas models, just had to be content with their simplest
isothermal treatment, saying that it was ``not a bad approximation'' overall
and, anyway, ``not significant for the linear mathematics from which we
obtained our main results.'' (\textit{Goldreich})} A thing to stop the growth and revert the
system to its initial state (else it is no machine) is to break the energy
replenishment of the gas layer. This done, it flattens and gets less stable.
Closer to marginal stability, the swing amplifier is turned on, it applies
to various existing perturbations,\footnote{ ``We felt that there were lots
of disturbances in galaxies once one mode had become nonlinear and so there
would be no difficulty in having a small component to amplify. We were not
concerned with any feedback loop at that time and to this day I am less than
sure of its existence in real unbarred galaxies as opposed to theorist's
models.'' (\textit{Lynden-Bell})} and analysis has it best tuned on those with $\lambda _{y}
\cong 8\pi h$, $h $ being the layer's half-thickness. In the nonlinear stage,
genuine trailing arms have been formed and hot stars are born in growing
condensations. They stir up the interstellar gas, however, and it swells,
recovering stability. The amplifier is turned off, the spiral arms break
down, the star formation stops, the hot stars fade away, the gas layer
thins, and the cycle repeats -- local structures on a scale $\lambda _{y}
\cong $ 1-2 kpc are periodically regenerated everywhere in the gas layer, the
only responsive galactic ingredient.

\bigskip
\bigskip
Now what to do with the disk of stars, another licensed player in galaxy
dynamics? Its natural length scale differs from that of the gas layer almost
exactly in the ratio of their column (surface) densities, or typically
roughly 10:1. At such a hostile difference their actual coupling cannot stop
the interstellar gas, well able to cool itself, from tending to have severe
gravitational instabilities of its own. Yet GLB reckoned that Jeans
instability ``occurs for stars in much the same way as it occurs for gas''
so that the ``spiral arm formation should [\ldots] be regarded [\ldots] as an
instability of the whole star-gas mixture''.\footnote{ The GLB gas treatment
of galactic disks reflected Lynden-Bell's earlier devotion to cosmogony and,
in its frames, to Ledoux-oriented analytical tradition of treating flat
systems (Ledoux 1951). GLB believed that even the largest-scale galaxy
dynamics features the gas as the colder and more pliable dynamical
component, and the idea of general `equivalent stability' of gas and star
models hit them on the fact that those obey the same Jeans-instability
criterion $\pi G\rho \ge \Omega ^{2}$ when infinite and in uniform rotation.
\par \textit{Lynden-Bell to Toomre}: ``We treated everything as gas not because
we think the gas is dominant (except possibly as a triggering mechanism) but
because in those cases where the transition from stability to instability can
be worked out for both a star distribution function and an equivalent gas system
they both become unstable at the same point. At present I only have a proof of
this for star clusters whose distribution functions depend on energy only and I
am not sure what equations of state the anisotropic pressure of a gas should
obey if it is to go unstable in the same way as the stars in the disk of a
galaxy. However I think this would probably make our basic philosophy clearer.
This work on the equivalence of stability is almost all that is directly relevant
that is happening here at present''. (Lynden-Bell 1964a) (``I had already found
very little difference between stars and gas in the Jeans instability criterion
so had little compunction in solving the gas problem with the velocity dispersion
of stars replaced by the sound velocity of the gas.'' (\textit{Lynden-Bell}))\par
\textit{Toomre to Lynden-Bell}: ``I understood that your main \textit{motive} was
not so much to gather what a supposedly smooth gas disk would do by itself, as to
mimic the likely behavior of a disk of \textit{stars}. At least in a vague,
intuitive sense I agree with you that the pressure should give neutral stability
results that should at the worst be of the correct order of magnitude. [\ldots
Still] the evident gross unevenness in the way the interstellar matter appears to
be distributed in most galaxies would have meant that such initially smooth
analyses could not \textit{directly} be relevant.'' (Toomre 1964c) \par
\textit{Lynden-Bell to Toomre}: ``I convinced myself that star and gas systems
(apart from static) normally have different critical stability criteria. This
floors my earlier hope though I did prove a nice theorem for static systems.''
(Lynden-Bell 1964c) ``When I get a typist to do it I will also send you a
lengthened version of the paper I delivered at IAU Symposium No 25 on stability
of collisionless systems. This is great fun though not applied to spiral problems.''
(Lynden-Bell 1964d)} Thus they coopted the star disk into their basic gas-dynamical
scheme and got a condominium with an essentially stellar `effective' density and --
clearly -- $\lambda _{T} $-comparable characteristic scale $\lambda _{y}
\cong 10$ kpc, ``embarrassingly large for something deduced from a
small-scale approximation''. ``From a local theory we cannot produce any
preference for the formation of symmetrical two-arm spirals'', GLB
recognized, but found it ``however [\ldots ] likely that the instability
leading to them is a somewhat more organized form of the one discussed
here'' (GLB, p.151).

\bigskip

The GLB paper was closed with a ``Note added in proof'' whose reproduction
here will allow us to turn conveniently to the subject of the remaining
sections of this chapter.

\begin{quotation} {\footnotesize
\noindent ``We have heard from Dr Toomre and Mr Julian of further work on zero
thickness stellar disks including a discussion of sheared modes. These
behave very similarly to their gaseous counterparts discussed here. This
work was independent of ours although the same sheared coordinates have been
invented by them.'' (GLB, p.157-158) \par}
\end{quotation}

\bigskip

\subsection*{1.4 Transient growth and asymptotic stability}
\addcontentsline{toc}{subsection}{1.4 \it Transient growth and asymptotic stability}

\bigskip

William Julian had, like Toomre, Kalnajs and Shu, been an undergraduate
student at MIT. After receiving his bachelor's degree in mathematics in
1961, he continued on as a graduate student and soon took a course on
galactic astronomy from Woltjer when he visited MIT. That roused Julian's
interest in galaxy dynamics, and the time, personified by Lin and Toomre,
magnified it. When the latter completed his axisymmetric-stability study of
flat stellar galaxies (Toomre 1964a), he determined to encompass the
asymmetric task, and this motif guided him and Julian into their work on
``Non-axisymmetric responses of differentially rotating disks of stars''
(Julian and Toomre 1966; hereinafter JT), which started in the spring of
1964. The news soon about parallel studies at the English Cambridge gave
them still more incentive to struggle along, upon which Toomre promptly and
in detail informed Lynden-Bell about the steps the MIT duet had done and
planned to do.\footnote{ Lynden-Bell and Toomre already knew each other.
They first met briefly in June 1962 at Woods Hole Oceanographic Institute.
Toomre's cold axisymmetric modal calculations were being finished during his
stay there, and he spoke of disk instabilities at a seminar with Lynden-Bell
present. (The listener later recalled: ``I fear that such are one's
subjective impressions that my memory of your talk at Woods Hole is solely
an irrelevance which I will not burden you with'' (Lynden-Bell 1964b). ``I
think your sentence is Churchillian'', then commented Toomre.) In June 1964
Mestel visited MIT, and he brought both Lin and Toomre preprints of two GLB
papers from Lynden-Bell. ``Figures 3-5 in their Paper (or really preprint)
II resembled hugely what Bill Julian and I had managed both to \textit{discover} and to plot
all on our own just during the preceding 1-2 months -- Toomre recalls. -- We
had at that point been doing our stellar dynamics only via truncated moment
equations which were flawed in not including the strong (= vaguely Landau)
damping toward short wavelengths that is very characteristic of the stellar
rather than gaseous problem ... and it was for that slightly bogus reason
that our results looked so similar.'' (\textit{Toomre})}

\begin{quotation} {\footnotesize
\noindent ``Since this May, a graduate student named W. Julian and I have been
involved in much the same sort of an analysis as you describe in your Part
II, but for the somewhat more complicated case of a thin sheet of stars with
not insignificant random motions in the plane of the disk. [\ldots ] My
interest in your problem dates back to the sheared non-axisymmetric
disturbances for the case of negligible pressure, which were among the
things I reported in the recent ApJ. Even at the time I did those, I
realized that any inclusion of pressure forces to remove the shortest
instabilities would leave a typical situation that was at first stable, when
the disturbance was still tightly wrapped in the `unnatural' sense, then
unstable for a while, and finally stable again. (In fact, if one were to
choose the unwrapped wavelength long enough, and the pressure quite small, I
felt one would even find two periods of temporary instability! Have you
tried this admittedly unrealistic case on your computer?)\footnote{ ``We did
not try any of the double growth period solutions (where oscillations take
place in between growths) because unless the radial modes are unstable the
double growth ones never get a decent acceleration.'' (Lynden-Bell 1964a)}
However, I felt then that the situation did not merit a detailed
calculation, since it could not be terribly relevant to the spiral problem
to discuss such disturbances to a supposedly uniform disk of \textit{gas} in view of the
observational evidence about the gross unevenness of the existing gas
distributions in galaxies. [\ldots ]

\bigskip
Certainly, you arrive at a most worthwhile result in observing that under
circumstances in which the axisymmetric instabilities (locally at least)
would be avoided, there is still the distinct possibility of temporary
non-axisymmetric instabilities, and that this could not help but provide a
bias in any situation with a somewhat random excitation in favor of waves
with the `natural' wrapped-up orientation. [\ldots ] Where I would at
present reserve my judgment is in your conclusion that your result is
directly pertinent to the spiral problem. Julian and I had our own burst of
enthusiasm on this when we obtained our very similar results, but lately it
has become a little more difficult for us to envisage the exact connections.
But surely it cannot be altogether irrelevant!'' (Toomre 1964b)\footnote{
``I agree with almost all you say, Lynden-Bell responded, even to some
extent the doubtfulness of whether the theory as outlined by us is really
the mechanism.'' (Lynden-Bell 1964a)} \par}
\end{quotation}

Using kinetic methods, Julian and Toomre described non-axisymmetric
responses in a thin Cartesian model of a small stellar-disk region of a
non-barred galaxy (JT model). In so doing, they actually managed to conquer
a considerably more difficult technical problem via the collisionless
Boltzmann equation than the one that GLB had needed to solve for their
idealized gas.\footnote{ ``My main idea in spring 1964 had been to expand
the perturbed phase density from the collisionless Boltzmann equation as a
sum of products of Hermite polynomials in $u $and $v$ multiplying the
two-dimensional unperturbed Schwarzschild distribution. Closing them was not
a big concern [... and] this was already some rather honest stellar
dynamics. [\ldots ] But Bill and I were dismayed to learn during summer 1964
(or roughly a month or two after the GLB preprints had arrived) that such
expansions looked as if they would need \textit{thousands} (!) of terms to begin to capture
reasonably accurately the later decay of vibrations due to what we realized
eventually was just phase-mixing. It was this terrible inefficiency that
prompted Bill to go looking extra hard at the alternative route of an
integral equation. And it was definitely he who first realized that the
ferocious kernel there could be integrated explicitly, an insight that
suddenly made that route \textit{much} more palatable than it had seemed at first''
(\textit{Toomre}). \par ``I guess Alar knew it also, but did not realize that the integrals
could be worked out.'' (\textit{Julian})} Help from the Volterra-type integral equation to
which the authors had converted the problem enabled them to track the
evolution of an impulsively applied disturbance and to see the shared waves\textit{ damping}.
They damped as well at a finite-time imposition of disturbances, while
asymptotically, as $t \to \infty $.\footnote{ That non-axisymmetric waves
damp is due to phase mixing of the perturbed star distribution function in
the course of its averaging. It does not imply energy dissipation as long as
the system obeys the isentropic collisionless Boltzmann equation. In the
gradient-free JT model, the mixing effect comes from the shear that breaks
down phase alignment of the stars on their epicycles, induced by previous
disturbances. Waves of lengths $\lambda _{y} > > 2\pi r_{e} $ ($r_{e} $
being the epicyclic radius) are almost uninfluenced, while those of $\lambda
_{y} \le 2\pi r_{e} \approx \lambda _{T} $ damp severely.\par ``Agris
Kalnajs started hammering away on my dense skull from roughly spring or
summer 1963 onward about the \textit{undamped} axisymmetric vibrations even in the presence
of ample (especially $Q$ > 1) random epicyclic motions, -- since they followed
very naturally (as he well knew) from the kinds of plasma-like math. [\ldots
] I remained suspicious for a long time especially about his claims that
there should be such undamped vibrations no matter how short one chose their
wavelengths -- I had somehow become over-convinced that strong phase mixing
or loosely speaking `Landau damping' of any short waves in stellar sheets
had to be the rule and could not be avoided! Of course on the latter point I
was wrong, as even Julian and I had convinced ourselves [...], \ldots but
intuition is a funny thing, and sometimes when wrong it takes a long time to
get repaired.'' (\textit{Toomre})}  ``This means, JT concluded, that a collisionless star
disk, if it strictly obeyed our model equations, could not even sustain
self-consistent non-axisymmetric waves set up by previous gravitational
disturbances, let alone admit modes that grow indefinitely'' (JT, p.819).
This plainly conflicted with the Lin-Shu self-sustained and tightly wrapped
wave scenery, while still giving it formally, as the axisymmetric limit, a
saving chance in an indefinitely slow damping.\footnote{ ``Let me state
again, a little more explicitly, why all that largely Russian `grumbling'
about folks in this business having been genuinely `ignorant of plasma
parallels' is beginning to get under my skin. In its linearized form, the
collisionless Boltzmann equation is nothing more than a 1st-order
quasi-linear PDE which almost any competent applied mathematician would (or
should) recognize is solvable via very standard characteristic curves such
as I did in my paper (Toomre 1964a). So I honestly still don't think that
was any big deal, or something that C.C. or anyone else could not quickly
rederive on their own. One most definitely did not need to go running to
plasma physicists to see how they had handled something so `obvious'.
[\ldots But] those characteristics surfaced again in JT, and there in a
situation with a shear flow which even the kind plasma physicists had
probably not met! I also assert that the Volterra integral equation (21)
from JT -- with its kernel figured out as compactly and explicitly as it
appears in eqn (23) thanks to my clever student Julian -- is distinctly
\textit{more} remarkable than anything Lin {\&} Shu 1966 managed to do on their own, esp.
since that formalism not only contains `their' dispersion relation as a
limiting case that we there only hinted at, but also because we unlike they
went on to show right in JT that the shearing sheet `could not even sustain
self-consistent non-axisymmetric waves', plainly contrary to what Lin {\&}
Shu 1966 would have implied for this same situation.'' (\textit{Toomre})}

\bigskip

These results enabled Julian and Toomre to speak of the stability in the
strict sense. Because the heat and shear parameters played no quantitative
role (as well as $k_{y} $, they were only demanded not to be infinitesimally
small), JT stated that the technically correct criterion for their model
disk had to be the axisymmetric one, $Q >$ 1. Yet ``this curiously simple
conclusion'' is but an asymptotic result, they stressed: ``as such, it does
not preclude the amplification of disturbances during a conceivable
intermediate time period'' (JT, p.819). And JT did compute ``a remarkable
transient growth of these wavelets while swinging'' (p.821), very similar to
that revealed by GLB in gas models.

\bigskip

\subsection*{1.5 Spiral stellar wakes}
\addcontentsline{toc}{subsection}{1.5 \it Spiral stellar wakes}

\bigskip

{\footnotesize \begin{list}{}{\leftmargin4cm}
\item The main inference to be drawn from these analyses is undoubtedly that even
a seemingly stable, differentially rotating star disk ought to \textit{respond} with a
remarkable intensity, and in a distinctly spiral manner, to quite typical
forms of non-axisymmetric forcing. [\ldots ] These intense trailing star
responses obviously demand a physical explanation.
\begin{flushright}
\textit{Julian {\&} Toomre 1966, pp.829, 827}
\end{flushright}

\item The response to a traveling point mass is a nice physical idea although
again it can't be very large scale. However I am sure one can see results of
it in galaxies and it could be important as an observational tool to tell
the conditions in a galaxy from the shapes and angles of `the tails of
condensations'.
\begin{flushright}
\textit{Lynden-Bell 1964d}
\end{flushright}
\end{list}}

\bigskip

\noindent However nice the above results from the first part of the JT paper may have
appeared, one cannot help noting that in the large they just confirmed the
basic GLB picture of the strong transient amplification. The `English
signal' that had come to Julian and Toomre in June 1964 must have made them
feel that they would not get out of the shadow cast by the GLB-planted
spreading tree without an advance in strict description of local
swing-amplification in its self-consistence and closure. And, it must be
said, that signal did not take them unawares: they had already set
themselves the task of answering the question: ``How would a thin,
differentially rotating, self-gravitating disk of stars respond to the
presence of a single, particle-like concentration of interstellar material
orbiting steadily within its plane?'' (JT, p.810)

\begin{quotation} {\footnotesize
\noindent ``We have thus far mainly talked about putting these disturbances together
to obtain among others the density patterns of the steady response of stars
to a point mass representing a similarly orbiting gas concentration, for
instance.\footnote{ ``It must have occurred to Bill and myself that the
forcing by a point mass was a basic question to be answered, since it
amounts essentially to a Green's function approach to this subject.''
(\textit{Toomre})} This task will only be messy, not difficult, [\ldots ] but we can
already foresee that the response density will be in the form of an
elongated hump, inclined roughly at your angle to the radius'' (Toomre
1964b).$^{} $ \par}
\end{quotation}

Volterra-equation methods allowed Julian and Toomre to calculate those
responses, essentially by superposing lots of individually shearing waves to
obtain a steady pattern of positive and negative disturbed densities in the
vicinity of the imposed mass point.~And precisely because many of those
waves had been strongly swing-amplified, this sum of Fourier harmonics
resulted in an awesome trailing stellar wake extending to both sides from
the point perturber (Fig.2). As might be expected, the isodensity line
inclination was sensitive to the shear rate and much less so to the
stability parameter $Q$; the latter, in its turn, strongly influenced the
wake's amplitude, especially at $Q \cong 1$. JT preferred $Q = 1.4$,
however, as corresponding to our solar vicinity, and for this case they
computed disk-thickness corrections. At the assumed thickness $2h \cong
0.1\lambda _{T} \cong 1$kpc, those reduced the perturbed gravity by no more
than 20-30{\%} but did not hurt the general characteristic picture of a
steady trailing spiral-shaped wake that impressed one with its severe length
scale and amplitude (Fig.3).

\begin{figure}
\centerline{\epsfxsize=1.0\textwidth\epsfbox{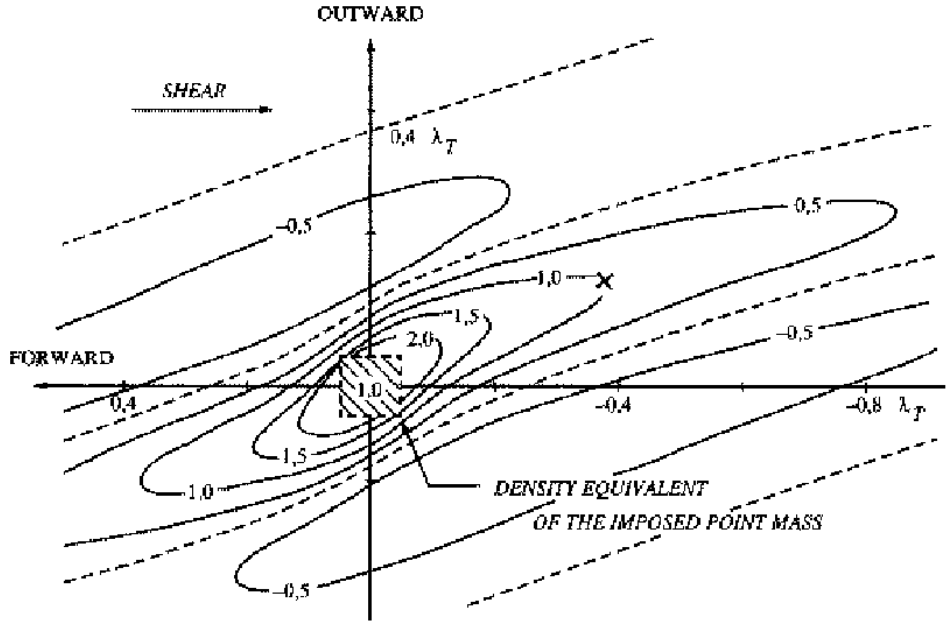}}
\caption{\footnotesize \textit{ A stationary density response of the JT model
on the action of a local mass source. $Q = 1.4$, $V(r) = const$.}
(The figure is reproduced from Julian \& Toomre 1966)}
\end{figure}

\begin{figure}
\centerline{\epsfxsize=1.0\textwidth\epsfbox{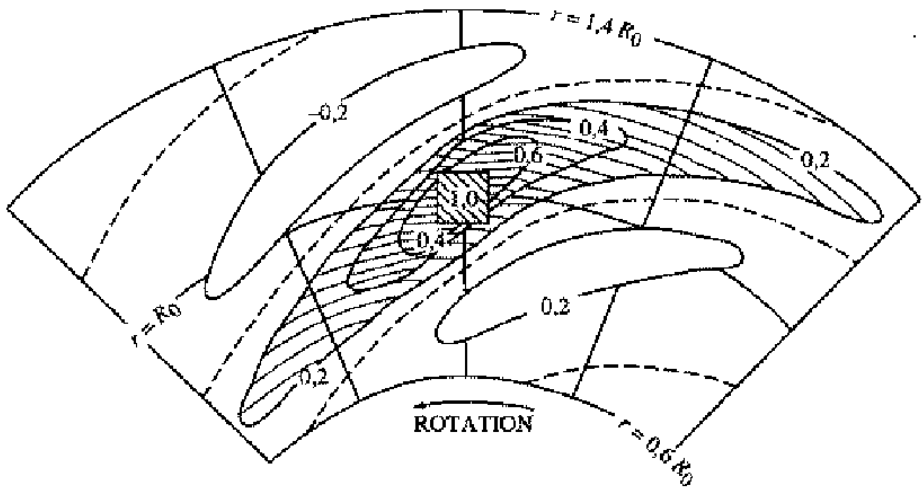}}
\caption{\footnotesize \textit{ A polar-coordinate view of the response in
Fig.2 corrected for the disk thickness $2h \simeq 0.1\lambda _{T}$.}
The galactocentric distance $R_0 $ equals $\lambda _{T}$.
(The figure is reproduced from Julian {\&} Toomre 1966)}
\end{figure}

\bigskip

To clarify the dynamical substance, the authors separately considered what
happens to a `cold' test star, say, on a larger circular galactic orbit than
the imposed point mass, as the differential rotation carries it past this
force center, collective forces being ignored (Fig.4). As long as the time
interval during which the star is close to the center is only a fraction of
an epicyclic period, as in the shearing conditions near the Sun, the radial
force component resembles an impulse occurring at the instant of closest
passage. It sets the star in epicyclic motion by giving it a radially inward
disturbance velocity at the abreast position. Because of this the relative
speed of passage reaches a minimum approximately one-quarter epicyclic
period later, or some 45$^{0}$ or so downstream of the perturbing mass
point. This angle, which is considerably larger in the collective case --
toward 70$^{0}$, as in Figs 3, 4 -- shows the direction in which the passing
stars are grouped most closely, forming a characteristic phase concentration
called a wake.

\begin{figure}
\centerline{\epsfxsize=1.0\textwidth\epsfbox{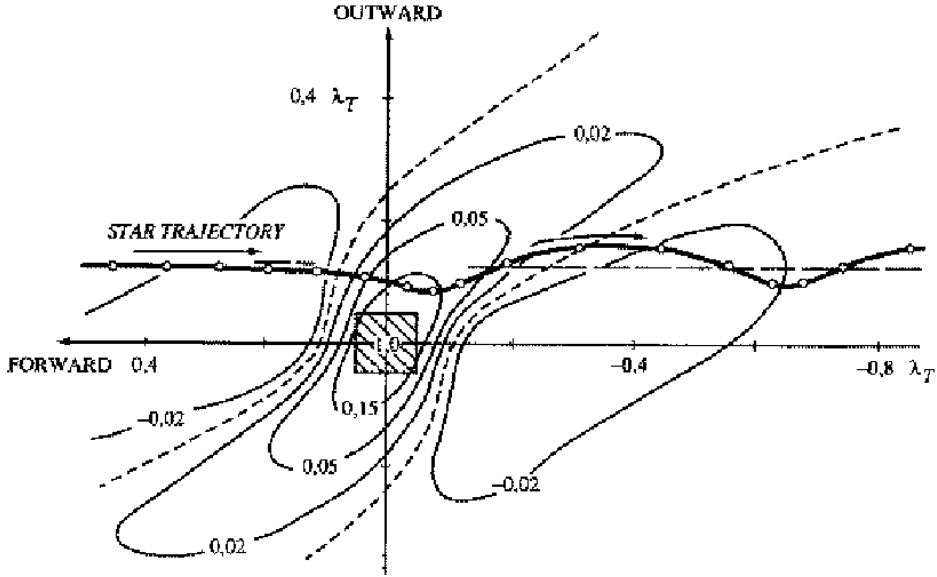}}
\caption{\footnotesize \textit{ Trajectory of a test star moving past a local mass source.}
$Q = 1.4$ and $V(r) = const$ as in Fig.2, but the collective
effects due to mutual attractions of the background stars are turned off.
(The figure is reproduced from Julian {\&} Toomre 1966)}
\end{figure}

\bigskip

\bigskip

What can one say about the JT work in summary?~It stands on its own as a
complete solution to a well-defined problem, an accurate and ample model, a
neat and strict theory (though tedious to compute).~Being self-contained, it
has no need for subsidiary assumptions, hypotheses, speculations and
evaluations.~It must have been evident to many thinkers that a steady
compact source might create nothing but a steady (what else, if any?) hump
of trailing (what other in the face of shear?) orientation.~Why had this
idea not been worked out earlier?~Because fresh physical intuition,
mathematical excellence and advanced computing were needed, and all at once.
But, all the same, the paper itself impeded general insight into its
findings.~Written with the feeling of intellectual and aesthetic pleasure of
having solved a difficult but important problem, the article contains some
unnecessary confusing details, and in other places -- through scrupulous and
otherwise brilliant style and wording -- is too condensed to be accessible
without a lot of work by the reader.~So it was rather too terse and
mathematical in the general climate of tastes and attitudes with which
traditional astronomers had encountered the early steps of new, modern
galaxy dynamics.~It has been no wonder that even several serious
dynamicists, let alone ordinary astronomers, have never bothered to read
this important paper carefully.\footnote{ The JT paper had fewer than 20
citations (according to ADS) in the first 10 years after its publication.}

\bigskip
\bigskip

\subsection*{1.6 The roads part}
\addcontentsline{toc}{subsection}{1.6 \it{The roads part}}

\bigskip

\bigskip

Tempted by a wealth of interests and prospects, the fathers of the `swirling
hotch-potch' theory were not very resolute about raising their
yet-unrisen-to-its-feet brainchild. Goldreich had no plans at all to
continue working on spiral structure, and once he had moved from Cambridge
to California in the summer of 1964 he did not pursue their studies. In the
late 1970s only did he return to the subject, and then only when he began to
study planetary ring dynamics with Tremaine (Goldreich {\&} Tremaine 1978,
1979, 1980).

\begin{quotation} {\footnotesize
\noindent ``You must appreciate that I was not a major player in the story you are
concerned with nor did I ever consider myself to be one. Moreover, I am not
a particularly scholarly scientist, and am undoubtedly guilty of paying too
little attention to who deserves credit and for what even on topics to which
I have contributed. My main pleasure comes from understanding things for
myself. I like to get applause for my work, but that is a secondary
benefit.'' (\textit{Goldreich}) \par}
\end{quotation}

No sooner had Lynden-Bell submitted the GLB article (spring 1964) than he
``just finished a paper (Lynden-Bell 1965a) on explaining the bending of the
galactic plane by a precession of the galaxy'' and then ``temporarily left
spiral research for a problem in general relativity'' (Lynden-Bell 1964a) --
apparently, not without a compunction on Goldreich's departure and partial
unsuccess of arranging things with Toomre so that they ``can work
complementarily rather than on the same topics'' (Lynden-Bell 1964c).$^{
}$``I quite expect to stop or rather remained stopped for a bit'' regarding
galaxy-disk problems, Lynden-Bell wrote to the latter (Lynden-Bell 1964d),
still in winter 1964-65 he collaborated with Ostriker on a general energy
principle for differentially rotating bodies. In his 1960 thesis
(Lynden-Bell 1960) he already had one for axisymmetric modes, and he was
eager now about the non-axisymmetric case envisaging its relevance to spiral
modes. The work was almost completed by the time Lynden-Bell left Cambridge,
and his summer 1965 arrival at the Royal Greenwich Observatory in
Herstmonceux returned him for a while to spiral regeneration channels.
Impressed by the ``formidable difficulties'' that the leading spiral
theories of the day met ``in fitting precisely the observed phenomena in our
Galaxy'', he announced that he was to show ``a fundamental role for a small
magnetic field in a basically gravitational theory'' via ``modification of
the Goldreich--Lynden-Bell--Toomre approach''. That, he believed, ``provides
a more natural discrimination between old stars and gas, avoids the
relaxation difficulties and provides condensations which do not spin too
rapidly for star formation.'' (Lynden-Bell 1966, p.57-58)

\bigskip

``I have gone over to being a magnetic man in part'', Lynden-Bell wrote to
Toomre (Lynden-Bell 1965b) inviting him for a bigger joint effort, which
struck the latter as a bizarre digression (\textit{Toomre}). Something must have been
disturbing him as to which spiral ideas to believe.\footnote{ ``I still do
not know whether the magnetism is an important catalyst for the modes we see
or whether it is irrelevant. It is irrelevant for a purely stellar system
but what we see has gas and star-formation.'' (\textit{Lynden-Bell})} $^{} $Possibly, this was
partly due to his general motif of finding pleasure in formulating and
trying original dynamical problems rather than in routinely clearing roads
already laid. There Lynden-Bell was no doubt successful since despite his
few miscarriages (like Lynden-Bell 1965a) he had won fame as a galaxy
dynamicist already in the 1960s, especially after his important studies on
violent relaxation of stellar systems (Lynden-Bell 1967) and on the nature
of quasars (Lynden-Bell 1969). With all that he himself has admitted, as if
implying the reverse of the medal:

\begin{quotation} {\footnotesize
\noindent ``I have no claim to the theory of spiral structure. Of those who once
worked on it I feel that I am one of those least well informed as to its
current state and most skeptical that a full understanding has even yet been
reached.'' (\textit{Lynden-Bell}) \par}
\end{quotation}

Working with Toomre on stellar wakes, Julian prepared his PhD thesis ``On
the Enhancement of the Random Velocities of Stars in Disk-like Galaxies'',
supervised officially by Lin and submitted in August 1965 (Julian
1965).\footnote{ ``Bill throughout our mutual involvement remained C.C.'s
student officially. [...] When I returned to MIT in fall 1963, C.C. himself
had urged me to look after Bill, not on the grounds that he wasn't talented
but because -- to C.C.'s own taste, at least -- he seemed too independent.''
(\textit{Toomre})} There and in his consequent paper (Julian 1967) he calculated the
heating of orbiting stars such wakes cause. The simple truth of local
differential rotation and triaxial residual-velocity ellipsoid had long
argued partial relaxation of our Galaxy's star disk but -- paradoxically --
found no reasonable explanation in terms of two-star encounters (see
Chandrasekhar 1942). In the early 1950s, Spitzer {\&} Schwarzschild (1951,
1953) proposed and qualitatively estimated the heating by giant `molecular
complexes'. Now Julian included collective star interactions and found much
higher growth rates and velocity-dispersion points: taking the `complex'
mass of an order of 10$^{6}$ -- 10$^{7} $ suns, he had Toomre's $Q-$parameter
grown to as large as 2.0 or so.\footnote{ Following Julian, Thorne (1968)
solved an inverse problem of dynamical friction on a massive particle in a
slightly eccentric orbit in a hot thin disk of stars. With JT techniques, he
included collective stellar interactions whose neglect had been excused for
pairwise stellar encounters in elliptical galaxies and galaxy clusters where
Jeans' length is of the order of the whole system, but not in flat galaxies
where it was co-ordered with their thickness, pointing at much more
pronounced collective effects. Thorne found that this collective play could
double the friction in magnitude.} This disfavored the Lin-Shu wave picture
ensured by the capabilities of marginally stable disks, yet no reasonable
reduction of Julian's generous choice for a typical gas-cloud mass was seen
to let $Q$ go under 1.4.

\bigskip

As a PhD-degree holder, Julian worked at the University of Chicago until
1967 when he got a postdoc position at Caltech. Goldreich warmly met him
there and had him running and swimming during lunch the second day already.
Soon they went on to do their famous work on pulsar electrodynamics
(Goldreich {\&} Julian 1969). For a while, Julian kept a side interest in
the continuing discussion between his distinguished former MIT colleagues,
but when Toomre wrote to him in 1970 musing about possible ``large-scale
sequels to the JT paper'', Julian -- now in New Mexico -- ``seemed to be not
at all interested'' (\textit{Toomre}).

\bigskip

\newpage

\section*{II. THE LIN-SHU THEORY GOES ON}
\addcontentsline{toc}{section}{II. The Lin-Shu theory goes on}

\bigskip

{\footnotesize \begin{list}{}{\leftmargin4cm}
\item If you believe that a spiral arm exists over a very large distance
in the Galaxy, you would probably also like to believe that it exists over
many rotation periods.
\begin{flushright}
\textit{Prendergast 1967, p.304}
\end{flushright}

\item In the beginning the immediate necessity was a consistent description
of the spiral phenomenon, in sufficiently good agreement with the observational
data.
\begin{flushright}
\textit{Bertin 1980, p.10}
\end{flushright}
\end{list}}

\bigskip

\noindent The Lin-Shu Milky-Way spiral diagram favorably met at the Noordwijk 1966 IAU
Symposium (Lin {\&} Shu 1967), its authors affirmed that their wave theory
already ``produced conclusions which appear satisfactory from a general
point of view''. It was declared ``free from the kinematical difficulty of
differential rotation''\footnote{ Woltjer, the key player who turned Lin to
galaxy dynamics, stated in his spiral review that the density-wave \textit{theory} as
pictured by Lindblad already ``resolves the kinematical difficulties, but of
course a dynamic justification is needed'' (Woltjer 1965, p.570). Lin
(1967b, p.458) soon claimed that his and Shu's theory resolves the same \textit{as well}.}
and permitting ``the existence of a [two-armed trailing] spiral pattern over
the whole disk while allowing the individual spiral arms to be broken and
fragmentary'' (Lin 1967b, pp.459, 462; Lin 1968, p.47). This optimism gave
Lin a feeling of confidence, correctness and leadership in the understanding
of galactic spiral phenomenon, feeding his further initiative.

\bigskip
\bigskip

\subsection*{2.1 Neutral modes and marginally stable disk}
\addcontentsline{toc}{subsection}{2.1 \it Neutral modes and marginally stable disk}

\bigskip

{\footnotesize \begin{list}{}{\leftmargin4cm}
\item ... a number of major improvements and further extensions of the theory.
\begin{flushright}
\textit{Shu 1968, p.5}
\end{flushright}

\item Still, a posteriori, the behavior of the system is remarkably simple, and
the use of asymptotics is a generous source of physical insight.
\begin{flushright}
\textit{Bertin {\&} Lin 1996, p.219}
\end{flushright}
\end{list}}

\bigskip

\noindent To Shu, his historic early coauthor, Lin posed the important task of
enriching the analytical attire of their `theory of density waves', and Shu
supplied some in his PhD thesis work ``The Dynamics and Large-Scale
Structure of Spiral Galaxies'', presented at Harvard in early 1968 (Shu
1968, hereinafter S68).\footnote{ ``Much of the credit for this
investigation belongs to Professor C.C. Lin who asked the key question
concerning the spiral structure of disk galaxies and then formulated the
basic approach toward the resolution of the problem. [\ldots ] Professor M.
Krook provided much generous help in his capacity as my faculty advisor and
official thesis supervisor. Without his guidance and patience, my progress
as a graduate student at Harvard would not have been as pleasant.
Discussions with Dr. A. Kalnajs have cast light on several major and subtle
points. Many of the more fruitful approaches were found only because of his
well-raised criticisms of the form of the theory prevailing at one time. I
have made use of some ideas of Professors A. Toomre and P. Vandervoort and
am indebted to them on that account. Professor Toomre's helpful criticisms
of various aspects of this research invariably proved to be illuminating.
[\ldots ] The arrangement and style [of the final draft of the manuscript]
were greatly improved by several suggestions made by Professor C.C. Lin. To
all of these people, I am extremely grateful''. (S68, p.i)} He started with
the derivation of a general integral equation for self-consistent responses
in a thin star disk. Kalnajs (1965) already had one in an epicyclic
approximation, attacking it for growing-mode solutions, but uninspired by
those arduous efforts, Shu was not in the mood to vie with him in direct
modal search, the more so as, following Lin, he targeted only tightly
wrapped neutral modes. This converted his practical interest in the integral
equation to analyzing its short-wavelength limit in the 2$^{nd}$ WKBJ order.
For the day, it was a rather worthy plan as for instance it seemed to allow
access to the radial behavior of the supposedly long-lived modes.

\begin{quotation} {\footnotesize
\noindent ``We start with the hypothesis that a neutral spiral density wave
exists. We then investigate the question whether such waves can be self-sustained
in the presence of differential rotation and finite velocity dispersion. In
this way, we are able to study, in a qualitative manner, the characteristics
of such self-sustained waves. This deals with the question of \textit{persistence}.
[...] We investigate the question how such waves can be expected to attain finite
amplitudes, and what mechanism is that allows them to take on a spiral rather than
a barred form. This deals with the question of \textit{origin}'' (S68, p.6) \par}
\end{quotation}

In his attempts of answering the so posed `question of persistence', Shu
resourcefully argued for adoption of the marginally stable galaxy-disk
model, and turning then to the `question of origin', he called for the idea
of overstability with which to resolve the `antispiral theorem' in favor of
a trailing quasi-stationary spiral mode.

\bigskip

Lin and Shu initially ascribed a quasi-stationary spiral structure to an
in-places-strongly-unstable star disk (Lin {\&} Shu 1964), but soon they
changed their mind (Lin and Shu 1966) for Toomre's early idea of the disk
entirely evolving to a state of marginal stability $Q = $1 (Toomre 1964a). Toomre
himself had already left it, having considered the role gas clouds must have
on stars, which Julian's calculations soon supported (Julian 1967), however
Lin and Shu remained skeptical of any need for $Q$ to rise above unity.

\begin{quotation} {\footnotesize
\noindent \textit{Lin:} ``Toomre (1964a) gave a criterion for the minimum
dispersion velocity needed to prevent gravitational collapse. He and Julian (JT)
are inclined to believe, however, that the mean square dispersion velocity might
exceed this minimum by as large a factor as 1.8. On the other hand, Lin and Shu
(1966) are inclined to believe that the value would not significantly exceed the
minimum needed [...] Since observations show deviations from a Schwarzschild
distribution, it is difficult to distinguish between these two opinions
without a careful analysis of the observational data.'' (Lin 1968, p.49)

\bigskip

\noindent \textit{Shu:} ``Whether the Galaxy is everywhere more than marginally stable is a point
of some debate. Julian (1967) is of the opinion that the enhancement of
cooperative effects of the irregular forces provided by massive objects (on
the order of $10^{6} - 10^{7}$ solar masses each) will inevitably drive $Q$ to
values substantially higher than unity. Observations in the plane of the
Galaxy show only the `spiral arms' to possess large mass concentrations.''
(Shu 1970c, p.111) ``In the density-wave theory of spiral structure (Lin and
Shu 1964, 1966), large aggregations of interstellar gas are the
manifestation of a density wave and do not represent either a bound or a
quasi-permanent body of matter. The interaction of stars with such a wave
does not lead to appreciable relaxation.'' (Shu 1969, p.506) \par}
\end{quotation}

This troublesome climate prompted Lin and Shu to reverse the logic of
thinking and they put, accordingly, that their pioneer Noordwijk plot best
attested its underlying $Q = 1$ star disk. Shu examined Lynden-Bell's
mechanism of violent relaxation and claimed it not occurring in disk
conditions. ``The only relaxation mechanism operative for stars in the early
life of such galaxies, he thus argued, is an axisymmetric form of the Jeans
instability discussed by Toomre'' (Shu 1969, p.505); it develops in the disk
plane and affects neither vertical distribution of stars nor their angular
momentum. Along the event sequence Shu proposed for this mechanism, our
young, still gaseous Galaxy first attains a disk form. Via shear
deformation, its mass distribution becomes axisymmetric, and turbulent gas
motions get fixed at a permanent level $c$ comparable to today's vertical
stellar velocity dispersion. There comes a period of violent star formation.
The baby stars, inheriting parental kinematics, gain an isotropic rms
velocity $c$. The fresh cold disk they arrange is a fit subject for the
operation of axisymmetric instability through which it heats up until a
stage $Q \equiv 1$ is reached. The process cannot go beyond it, and losing
the heat is also impossible owing to the lack of any plausible cooling of
the stars (Shu 1968, 1969, 1970c).

\bigskip

In the adoption of $Q \equiv 1$ Lin and Shu found two attractive factors.
One was that in this neutral-mode case the four dispersion-curve branches
seemed to converge at corotation $\nu = 0$ (Fig.5). Shu conceived that two
longer-wave branches, due mainly to differential rotation, are ``more in the
nature of pulsations'', and two other, determined primarily by velocity
dispersion, are ``more in the nature of \textit{local} oscillations'' (S68, p.108). Still,
well seeing that these two processes are present in varying degrees here and
there in the disk, he found this ``useful for conceptual purposes'' insight
``somewhat arbitrary'' and credited realistic `coherent' spirals ``without a
`kink' at $\nu = 0$'' to a proposed `Mode-A' meant to couple the short-wave
branch inside corotation with its long-wave counterpart outside the
same.\footnote{ Shu's proposed `Mode-B' combined the long waves inside and
short waves outside corotation. ``Formally, Mode-B spirals with $m$ = 2 would
present the appearance of a barred spiral'' (S68, p.123).}  This smooth and
conscious selection, Shu noticed, had already served him and Lin in 1966
with their Noordwijk Milky-Way spiral understood as the inner half of
Mode-A.

\bigskip

Lin and Shu hoped that ``after a galaxy has been completely stabilized
against Jeans' condensational instability, it is still susceptible to a mild
overstability of two-armed waves'' to which one owes actual spiral formation
(S68, p.8).\footnote{ Overstability meant to Lin and Shu (Lin {\&} Shu 1966;
Lin 1967a) slow growth of waves traveling in an inhomogeneous $Q = 1$ disk.}
Shu developed this theme in his thesis. In 1967 he learned from Toomre about
his tentative group-velocity results and misused those to visualize how the
individual \textit{wave crests} move radially. He did not think then of genuine spiral-wave
packets (see Sect. 3.2) and what he had was but a group of tightly wrapped
two-armed waves somehow occurring to a galaxy and soon developing into an
almost self-sustained mode, to get perfectly so via slight shearing and
other modifications when it would gain and fix its amplitude. But if such a
wave-crest group is not quite a mode yet, why not to apply to it
group-velocity formulas? Shu did so and there he saw ``another (and perhaps
more important)'' attraction due to the $Q = 1$ model (S68, p.111): his
near-Mode-A got an inward radial group motion that ``does not reverse sign
somewhere in the principal range'' between the inner and outer Lindblad
resonances (ILR and OLR hereinafter).\footnote{ It is this mode, Lin and Shu
believed, that manifests itself in the observed spiral structure, and only
by superposing the identical `near-mode' with opposite sense of winding and
direction of motion that ``we obtain pure standing waves which do not
propagate. Such a wave, of course, does not have any spiral features.''
(S68, p.113)} In the inhomogeneous overstable disk such a motion ``would
lead to the growth of a `group' of spiral waves to some finite amplitude,
the growth being ultimately limited by non-linear effects [\ldots of] the
shearing effects of differential rotation (which is absent in the linear
theory) [that] may be expected to enhance any preference for trailing
patterns.'' (p.8)\footnote{ Lin and Shu knew well that the shear, which was
absent in their wave-mode theory, was absolutely present in the alternative,
sheared-wave theory (GLB, JT) and that it there supported nothing but
trailing waves. At the time they (and not only they) thought, however, that
there was no intrinsic connection between these two types of density-wave
theories. From their own end, they were not very successful in the 1960s in
explaining the trailing-spiral prevalence, though that had been a vital test
for any spiral theory. As regards their repeated mentions of and hints at
nonlinear effects (Lin {\&} Shu 1966; Lin 1967b, 1968), Lin and Shu never
went into it very seriously. Besides, their view of mild instability favored
trailing waves only inside corotation, diagnosing that ``there might be a
preference of leading waves in the [Galaxy's] range 10-12 kpc''. Lin (1967a,
p.80-81) professed that this ``cannot be taken on face value'', and largely
to avoid the trouble he left his original grand spiral plan over the entire
`principal region' between the $m =$ 2 ILR and OLR for a conceptually different
but yet space-preserving variant with corotation just transplanted to safer
areas of disk outskirts or thereabouts.\par Soon Contopoulos (1970a,b)
calculated near-ILR stellar orbits subjected to a growing imposed Lin-Shu
spiral gravity field of leading and trailing planforms. He got trailing
responses in both cases and explained his result in terms of a specific
character of misalignment of solutions inside and outside ILR, considering
this the first strict demonstration of the Lin-Shu-wave trailing.}

\begin{figure}
\centerline{\epsfxsize=0.9\textwidth\epsfbox{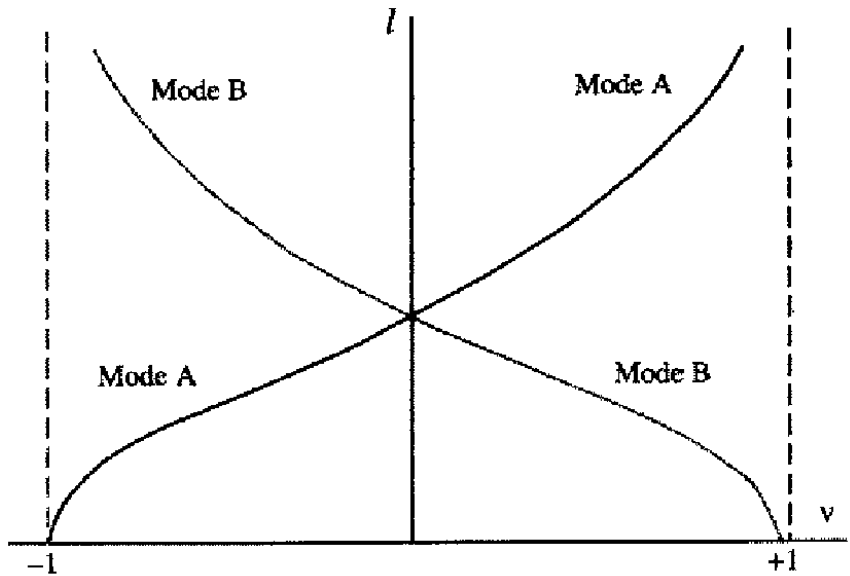}}
\caption{\footnotesize \textit{ Shu's Mode-A as proposed by him to account for
the grand design in non-barred spiral galaxies.}
(The figure combines two separate figures from Shu 1968)}
\end{figure}

\bigskip

These overstability ideas Shu directly associated with the \textit{antispiral theorem} that ``a number
of us have sometimes been worried'' about (Prendergast 1967, p.308).

\bigskip

\subsection*{2.2 Antispiral theorem}
\addcontentsline{toc}{subsection}{2.2 \it Antispiral theorem}

\bigskip

{\footnotesize \begin{list}{}{\leftmargin4cm}
\item After my paper on the stability of collisionless gravitating spheres was
published (Antonov 1960), ${} $ I approached the density-wave theory but did not
believe it, mainly because of the antispiral theorem, anyway known to
physicists.
\begin{flushright}
\textit{Antonov 2003}
\end{flushright}

\item In hindsight, I think Lin's judgment was accurate considering how quick
people were to attack his point of view with proofs of `antispiral theorems'
and the like shortly after the publication of Lin {\&} Shu 1964.
\begin{flushright}
\textit{Shu 2001}
\end{flushright}
\end{list}}

\bigskip

\noindent At the Noordwijk Symposium Prendergast explained to astronomers the general
meaning of the theorem. If in linear theory there were to exist a
nondissipative global mode of trailing planform that was content to rotate
indefinitely without growing or decaying, then a similar mirror-image
leading mode must exist as well. ``This symmetry property of the equations
means only one thing: the system is too simple. Whenever you see a symmetry
property, all you have to do is mess up the system a little bit and give up
the symmetry. There are a large number of things that will remove the
symmetry, [\ldots ] there is non-conservation of everything'' (Prendergast
1967, p.308-309).\footnote{ The feel of symmetry breaking ``non-conservation
of everything'' then prompted Prendergast that there ought to be some way to
determine that ``the natural way to get the arms is trailing'' and that
``presumably that would be a direction that would be given [\ldots ] by an
increase of entropy'' (Prendergast 1967, p.309).}

\bigskip

The antispiral theorem took on particular sounding after Lynden-Bell and
Ostriker (1967) set it out as an application of general principles they
worked out for differentially rotating bodies. Lynden-Bell, to whom we owe
the idea of this explicit consideration, no doubt knew that it ``had many
let-outs'' hence he ``did not think it as restrictive of spiral theories as
some others took it to be''. For one thing, the theorem could be strictly
applied to exponential modes only, and Lynden-Bell hoped that ``double modes
that might grow as $t\exp (i\omega t)$ might well be the ones needed to
transfer angular momentum outward'' through corotation
(\textit{Lynden-Bell}).\footnote{ ``I always held the view that angular-momentum transfer is the
driving force behind spiral structure. [\ldots ] In part the anti-spiral
theorem was there because it seemed to point out that what Lin said was much
less than the whole story.'' (\textit{Lynden-Bell})} Moreover, it did not oblige one at all to mix
leading and trailing waves in equal proportion obtaining a cartwheel-type
mode, that was no necessity imposed by the equal-frequency condition. Shu
recalls that Lin from the beginning ``felt sure that one should not do the
na\"{\i}ve thing of superimposing equal trailing and leading parts'' and
that ``he probably wanted to discover the reason why before publishing
anything'', but the Toomre 1964a paper ``triggered him into premature
action'' (\textit{Shu}). One is to wonder what annoyance for Lin and Shu became
Lynden-Bell and Ostriker's antispiral address that appeared just when they
thought they got the true mixing mechanism as due to disk overstability. It
was imagined to cause slow growth of one of the components, the trailing one
in Lin-Shu's `nonlinear' assumption, and then to break the full symmetry in
the basically neutral-mode problem by ensuring different radial behavior for
the components and, correspondingly and automatically, their unequal mixing.

\bigskip

In his 1968 thesis and, more pointedly, in his papers to follow (Shu
1970a,b) Shu demonstrated one more `let-out' in the antispiral theorem.
``The general formulation for the normal modes, he noticed, [\ldots ] shows
that a certain degree of spiral structure must be present in every mode of
oscillation which contain stars in resonance'' (S68, p.7). Stars, unlike
gas, can resonate with the oscillating gravity field without any continual
shattering due to collisions. Mathematically, this is answered by the
integrand poles, and even at real frequencies those compel one to make
integrations along contours going off the real axis, which provides the
solutions with an imaginary part and ensures their general spiral form. The
resonant technique of clearing the antispiral hurdle was to Lin and Shu one
of the highest points to back up the QSSS as a neutral density wave (Shu
1970a,b; Lin {\&} Shu 1971).\footnote{ Kalnajs already in 1963 had an idea
of such a resonance `resolution' of the antispiral theorem in the
neutral-wave setting (Kalnajs 1963; see Paper I, Sect. 2.4).}  It seems
curious, however, that they did not refer to any leading component either in
1966 on their short-wavelength spiral proposal for our Galaxy (Lin {\&} Shu
1967), or in 1971 when Shu et al (1971) announced for л51 and л81, apart
from their dominant short trailing waves, unmistakable traces of an extra
`mode', yet not mirror-reflected -- short and leading -- but \textit{long} and again
trailing.

\bigskip

\subsection*{2.3 Spiral shock waves and induced star formation}
\addcontentsline{toc}{subsection}{2.3 \it Spiral shock waves and induced star formation}

\bigskip

{\footnotesize \begin{list}{}{\leftmargin4cm}
\item Fujimoto, ${} $ followed by Lin and Roberts, ${} $ recognized that gaseous motions
generated by a tightly wrapped density wave would be dominated by the
appearance of tightly wrapped shock waves. Later work (Shu et al 1972) has
fulfilled Lin's belief that the density wave itself might trigger star
formation and it is the shock that seems to be the trigger.
\begin{flushright}
\textit{Lynden-Bell 1974, p.117}
\end{flushright}

\item It is not astonishing that one gets difficulties in making stars. I think
nature has difficulties too, because otherwise no interstellar matter would
be left.
\begin{flushright}
\textit{Hoerner 1962, p.107}
\end{flushright}
\end{list}}

\noindent ``In the early 1960s, Prendergast often expressed the view that the intense,
slightly curving dust lanes seen within the \textit{bars} of such SB galaxies as NGC 1300
and 5383 are probably the result of shocks in their contained gas, which he
believed to be circulating in very elongated orbits. In such `geostrophic'
flows of presumed interstellar clouds with random motions, Prendergast
(1962, p.220) wrote that ``it is not clear what is to be taken for an
equation of state'', but he knew that ``we should expect a shock wave to
intervene before the solution becomes multivalued''. As regards the normal
spirals, Lin and Shu (1964) stressed from the start that since the gas has
relatively little pressure, its density contrast ``may therefore be expected
to be far larger than that in the stellar components'' when exposed to a
spiral force field such as they had just postulated. That hint remained
largely dormant, however, until Fujimoto [...] combined these last two
lines of thought'' (Toomre 1977, p.453).\footnote{ As such, the idea of a
shock in the interstellar gas was not novel in the mid-1960s as for years it
had helped various small-scale problems. So the larger-scale shock-wave
speculation did not come to be very striking. Goldreich and Lynden-Bell
(1965a,b), for instance, had it quietly on their spiral-regeneration
concept. In fall 1964 Lynden-Bell remarked in his letter to Toomre, musing
on the topics for their intended complementary work: ``The other thing that
is interesting is the formation of shock waves as the disturbances get very
violently sheared, but while the overall structure of spiral arm formation
is not very clear I would rather leave such a secondary problem till later''
(Lynden-Bell 1964c). ``I profess no vested interest in the formation of
shock waves in an initially smooth gas layer, much against my upbringing as
a fluid dynamicist, Toomre responded. All yours.'' (Toomre 1964d) ``I am not
going to work in shock waves for a few years yet, came Lynden-Bell's upright
reaction, so you are welcome to them too.'' (Lynden-Bell 1964d)}

\begin{figure}
\centerline{\epsfxsize=\textwidth\epsfbox{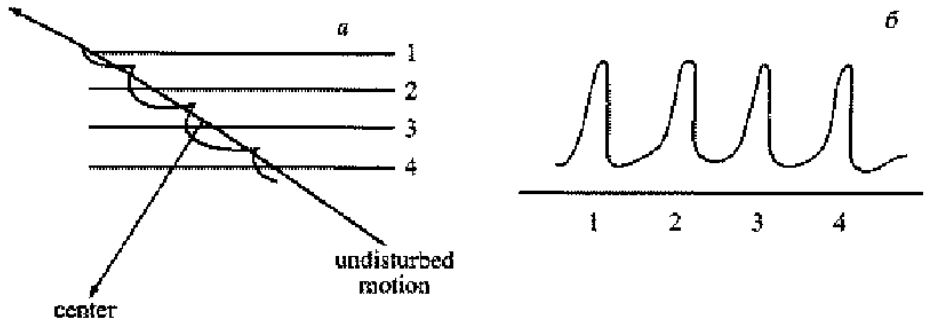}}
\caption{\footnotesize \textit{ The behavior of interstellar gas in the spiral
gravitational field of a galaxy, according to Fujimoto:
(a) -- the gas motion across the ``gravitational washboard'', (b) -- the
ensuing gas density distribution.} (The figure is reproduced from Prendergast
1967)}
\end{figure}

\bigskip

Fujimoto made his spiral-shock-wave report at the May 1966 IAU Symposium
held in Burakan, Armenia, but it became widely known thanks to
Prendergast.\footnote{ The Symposium proceedings were published in Russian
2.5 years later.} He not only had urged Fujimoto to consider the problem and
``helped much via fruitful discussion'' (Fujimoto 1968, p.463), but also
presented his results at Noordwijk, just three months after Burakan.

\begin{quotation} {\footnotesize
\noindent ``Let us suppose one has a rotating system and a \textit{gravitational washboard}, that is, a disturbance in
the gravitational potential of a sinusoidal form. [...] Then what
happens to the gas? [...] The answer is that the streamlines of the gas
-- instead of being straight lines, in this model corresponding to a perfect
circular orbit -- become somewhat cusped. In every cusp there is a shock;
and in that shock the density increases, even for a very modest
gravitational washboard, by an enormous factor; let us say five or more''
(Prendergast 1967, p.310). \par}
\end{quotation}

Fujimoto confirmed this shock-wave picture (Fig.6) by computing
two-dimensional nonlinear dynamics of perfect isothermal gas in a
quasi-steady field of galactic spiral potential. As the spiral angular speed
``cannot be determined even by Lin's method'', he decided to ``take a priori
some reasonable values'' and chose, as Lindblad (1963) and Kalnajs (1965)
did it, high speeds $\Omega _{p} \cong 45$ km/sec/kpc and close corotation
$r_{c} \cong 5$ kpc. The answer he gave was that ``both the high-density
hydrogen gas contained in spiral arms and the dark lanes seen in external
galaxies on the concave side of their bright arms can be due to the presence
of the shock waves'' (Fujimoto 1968, p.463). Lin reacted quickly and
inspiringly. He assumed that a shock wave could cause and
organize star formation in the spiral arms (Lin 1967b), and posed as William
Roberts' thesis theme the problem of modeling ``the presence of large-scale
`galactic shocks' that would be capable of triggering star formation in such
narrow spiral strips over the disk'' (Roberts 1969, p.124].

\bigskip

Roberts considerably developed and expanded Fujimoto's analyses. He
corrected one mistake made by Fujimoto with his working equations (he had
missed one of the full-value terms in the perturbed gas velocity equations),
presented the star-disk potential description in the Lin-Shu asymptotic
language, and focused on slowly rotating two-armed spirals with distant
corotation. Roberts' interest was in a ``particular type of solution of the
nonlinear gas flow equations'' permitting gas to pass through the shock
waves coincident with spiral equipotential curves and describing the gas
flow along a nearly concentric closed streamtube band, to exclude net radial
transfer of anything. And indeed he got desirable solutions whose family
presented ``the composite gas flow picture over the whole galactic disk''
(Roberts 1969, p.129). To this he conjectured that the shock wave, unaided
by large-scale magnetic fields (which were Lin's initial candidate (Lin
1967b)), could trigger by itself the along-spiral formation of star
associations.

\begin{quotation} {\footnotesize
\noindent ``One might imagine that the gas in turbulent motion has `clouds' before the
shock, which are on the verge of gravitational collapse; the sudden
compression would then trigger off the collapse of the clouds, which would
lead to star formation. After the gas left the shock region, it would again
be decompressed, and the process of star formation would cease'' (Lin et al
1969, p.737). \par}
\end{quotation}

\begin{figure}
\centerline{\epsfxsize=0.7\textwidth\epsfbox{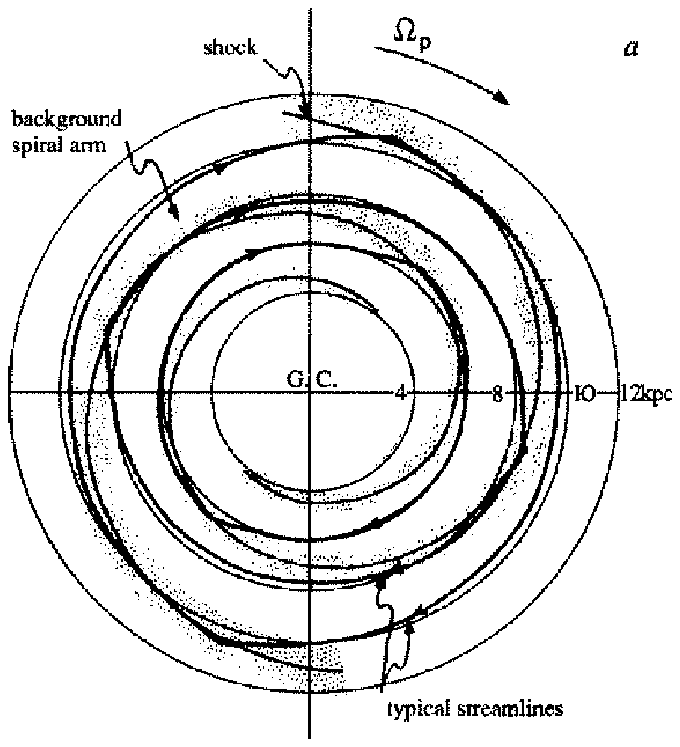}}
\centerline{\epsfxsize=0.7\textwidth\epsfbox{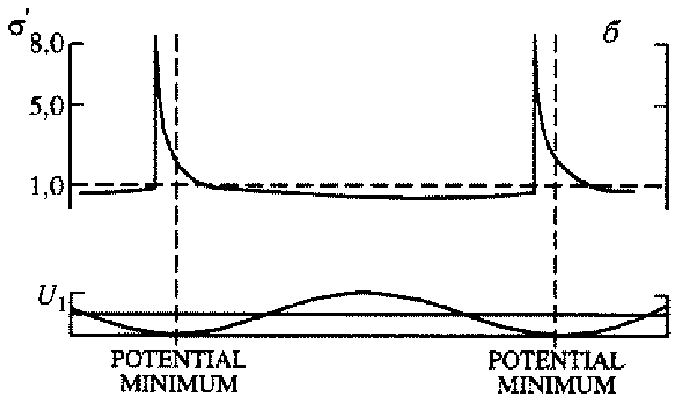}}
\caption{\footnotesize \textit{The behavior of interstellar gas in the spiral
gravitational field of a galaxy, according to Roberts:
(a) -- a shock in the gas as its reaction to the spiral gravitational
field of the stellar disk component, (b) -- azimuthal density
distributions in the stellar disk and interstellar gas.}
(The figures are reproduced from Roberts 1969)}
\end{figure}

How could this sudden growth of the interstellar gas density and pressure
trigger the desired gravitational collapse of the already existing dense
clouds? Roberts did not know or show -- he only said: conceivably (Roberts
1969, p.131) -- but that hardly matters.\footnote{ ``However, the previous
discussions (Roberts 1969; Lin et al 1969) are incomplete -- Shu and Roberts
admitted. -- There are severe difficulties in visualizing how this
`effective pressure' is transmitted on a small scale to trigger the
gravitational collapse of clouds since cloud-cloud collisions provide
compression essentially only in one direction''. Looking for ``a clear
physical basis for the mechanisms of the production of the shock and of the
compression of the clouds'', they discussed a two-phase model of the
interstellar gas (Shu et al 1972, pp. 558, 585).}  ``The crucial point is
that before the shock idea there had been no defensible explanation at all
for the striking geometrical fact, first noticed by Baade in the late 1940s,
that the main HII regions in large spirals tend to define considerably
crisper and narrower arms than the rest of the visible material [\ldots and
that] these highly luminous chains seem biased toward the inner edges of
arms, though perhaps not quite as much'' (Toomre 1977, p.453-54).\footnote{
``Two important assumptions underlie the above [Roberts'] gas dynamical
results. The first is that the driving potential wave is tightly wound and
the second that the interstellar medium can be adequately described as an
isothermal gas. It is a great pity that these were introduced in the initial
stages, since the results in many minds have been tightly associated with
them. However, most of the results still stand if these assumptions are
relaxed, e.g. it has been shown a number of times that a very open or even
barred forcing can drive a similar response.'' (Athanassoula 1984, p.348)}

\bigskip

Roberts' 1969 work was greatly appreciated as a basis for further studies of
related problems in galaxy physics. Well familiar to astronomers in various
contexts, the shock-wave idea in the new, spiral context appeared to many to
be more attractive and soluble than the intricate collective effects from
collisionless star-disk dynamics.

\bigskip

But there was some deserved criticism as well, largely in relation to
Roberts' stress on closed gas streamlines. Pikelner (1970), who calculated
energy loss of the gas as it crosses a spiral-wave arm and found it to be a
few percent of its kinetic energy, concluded that in order that the gas
might gradually come closer to center its streamlines had to be
open.\footnote{ The latter factor, Pikelner noticed, ``must have the
cosmogonical consequence, [\ldots ] since in the lifetime of an Sc-galaxy
gas must cross its [every] arm dozens of times. [\ldots ] Possibly, this
explains why in spite of the intense star formation in galactic innermost
parts there still remains a fair amount of gas.'' (Pikelner 1970, p.758).}
He noticed also that ``gas flows out of the arm slower than at the
reversible compression, so that its mass center shifts behind the arm axis
and pulls the stars ahead'' (Pikelner 1970, p.758). Therefore, he inferred,
angular momentum transfer from the shock-suffered gas to the spiral wave
must amplify the latter, feeding it up with energy. Kalnajs corrected
Pikelner in absentia (Simonson 1970; Kalnajs 1972).\footnote{ Kalnajs
advanced his criticism in the course of a discussion at a special `Spiral
Seminar' held in 1970 at the University of Maryland in connection with the
problems of spiral structure as followed from the findings of the recent IAU
Symposium at Basel, 1969 (see Simonson 1970). After, he discussed the
subject corresponding with Roberts and then set it out in his special note
(Kalnajs 1972). Shu and Roberts rejoined (Shu et al 1972; Roberts {\&} Shu
1972). They agreed that, strictly speaking, gas streamlines could not be
taken closed, but emphasized that in the WKBJ limit the actual non-closure
per cycle was a quantity proportional to small spiral pitch angle, which
Roberts reasonably neglected in his original paper. It must be said, however,
that originally he was rather straightforward about following Lin's directive on
a \textit{stationary} picture (Lin 1967b, p.463), and in that way he even created
some photogenic theory of gas `free modes' (Roberts 1969), which was really more
than just asymptotic.\par ``Yup, these `free modes' happily lose energy, and
keep on doing it forever without suffering any damage. This discovery deserves
to be commercialized. [\ldots ] What is the secret that makes the streamlines
closed? According to Appendix V [in the PhD thesis by Roberts (1968)], you just
assume that in the equilibrium state there are axisymmetric radial \textit{and}
tangential forces.'' (\textit{Kalnajs}) \par For more detail on this discussion see
Toomre 1977.} $^{} $Having established that the Lin-Shu slow waves carry
over \textit{negative} energy and momentum (Kalnajs 1971; Lynden-Bell {\&} Kalnajs 1972), he
recognized that their `pulling ahead' should cause the \textit{inverse} effect, i.e. wave
damping. All in all, the conclusion was that in open spirals (and bars,
providing the general mass distribution in the SB-galaxies is also roughly
axisymmetric) the shock wave ``still more increases density and
non-reversible dissipation of energy'' (Pikelner 1970, p.758). ``It meant
not only that Roberts was slightly mistaken [\ldots ]. Much more important,
this [\ldots ] implies that even the neatest spiral structures can at best
be only quasi-steady'' (Toomre 1977, p.460).

\bigskip

\subsection*{2.4 Extremely satisfactory comparisons?}
\addcontentsline{toc}{subsection}{2.4 \it Extremely satisfactory comparisons?}

\bigskip

{\footnotesize \begin{list}{}{\leftmargin4cm}
\item \textit{J.H. Oort} asks: What are Lin's further plans for numerical model computations [of
spiral structure]?

\item \textit{Lin} answers: At present, we have no immediate plans to extend our work much
further\ldots In the meantime, we are collaborating with Stromgren on the
problem of the migration of stars, to gain a more definite picture of the
spiral structure within a few kpc of the Sun.
\begin{flushright}
\textit{Discussion: Noordwijk 1966, p.334}
\end{flushright}
\end{list}}

\bigskip
\noindent Being in 1966 on the wave of his first success in astronomy, Lin
spoke of ``three levels of discussion in dealing with the structure of the
Theory of Spiral Structure'' (Lin 1966b, p.6). Two of them -- physical and
mathematical -- he saw already mastered to a degree,\footnote{ All the same,
Lin conceded that ``as one can see upon a little reflection, the problem of
the origin of the spiral structure is mathematically more difficult'', so
that ``these studies remain a challenge for future investigations'' (Lin et
al 1969, p.722).} so that more opportune and vital for the day he recognized
the third level that discussed ``the agreement of the theory with detailed
checks with observations'' (Lin 1966b, p.6).

\begin{quotation} {\footnotesize
\noindent ``In view of the difficulty of the theoretical problem, it is fortunate
that, from the beginning, we have placed great emphasis on working out the
\textit{consequences} of the QSSS hypothesis'' (Lin 1975, p.120). ``In the absence of a complete
theory for the mechanism of density waves, the need for observational
support is urgent'' (Lin et al 1969, p.722). ``Indeed, our theory can be
used as a tool to connect several seemingly unrelated observations. [...]
Even without the discussion of the detailed mechanisms, the mere assertion
of the existence of a density wave with a spiral structure, propagating
around the galactic center, leads to implications which can be checked
against observations.'' (Lin 1968, p.36, 49) \par}
\end{quotation}

Deliberately avoiding mathematical difficulties as ``the challenge for
future investigations'', Lin did not seem embarrassed in the face of
empirical difficulties due to apparent incompleteness and inaccuracy of
observational data, and he went and led his associates along the way that
was in fact no lesser challenge. There he saw an urgent interest in problems
of systematic noncircular gas motions and star migration (Lin 1966b; 1967b;
1968). The famous 1969 paper by Lin, Yuan and Shu ``On the structure of disk
galaxies. III. Comparison with observations'' (Lin et al 1969, hereinafter
LYS) gave a summary of all of Lin's ``levels of discussion''.

\bigskip

Systematic noncircular gas motions in the Galaxy were under discussion
already,\footnote{ Kerr (1962) was likely the first to point out systematic
motions as a possible source for differences in the northern and southern
observations. Considering such motions near the outer edge of the
Sagittarius arm, Burton (1966) suggested that an along-arm hydrogen flow
might explain the high-velocity stream. Shane and Bieger-Smith (1966)
considerably contributed to the discussion.}  still -- LYS noticed -- no one
spoke of their producing dynamical mechanism, although ``it is easy to see
this from considerations of angular momentum''. Indeed, the spiral gravity
causes additional along-arm gas motion which must be with the general
rotation on the outside edge of the arm and against it on the inside edge,
and which is ``to maintain \textit{conservation of matter}'' (LYS, p.731). Starting from the well-known
data on wavelike velocity variations in the Galaxy rotation curve,\footnote{
``These variations have long been observed, but they were thought to be
possibly the consequence of missing gas over interarm regions. A detailed
study by Yuan (1969a) has conclusively shown that the latter effect does not
give significant contributions to the variation in velocity.'' (LYS, p.731)}
LYS determined radial and azimuthal components of noncircular motion to be
of the desired, for linear analysis, order of 10 km/sec, and assured (Lin
1966b; Yuan 1970) that the observed three-to-five-fold gas compression in
the arms just corresponds to such a motion, so that ``there is good
agreement in major features'' (LYS, p.732). But at their accepted pitch
angle $i \cong 5^{0}$ WKBJ equations associated those motions with a rather
strong, knowingly nonlinear density response that was the business of a
theory yet not in the authors' hands. Thus more correct would be their
simpler inference that the observed noncircular motions indeed ``in major
features'' are due to a \textit{certain} spiral density wave.

\bigskip

Mentioned by Lin at Noordwijk, the problem of young star migration arose in
relation to Stromgren-initiated studies of star ages, claimed to be accurate
within 15{\%} (Crawford {\&} Stromgren 1966; Stromgren 1966a; 1967)].
Contopoulos and Stromgren (1965) set this migration problem as based on the
idea that time reversion of star motions and their countdown in the hold of
the axisymmetric gravity field of the Galaxy could / would show the stars'
birthplaces and check if they fall inside the arms. With their tables of
plane galactic orbits, Stromgren traced back the migration history of about
sixty late B stars aged between 100 and 200 million years and placed within
200 pc from our Sun. Those ``showed a definite separation into two
[velocity] groups'', and their birthplaces took a nearly tangent-to-circle
extended area connecting those regions of two nearest to us outer arms where
the `points' were grouped more closely. Stromgren concluded that ``the
present location of the arms favors the picture formed by the theory of
density waves, providing one takes the pattern frequency $\Omega _{p} $ to
be about 20 km/sec/kpc'' and that this ``offers possibilities of testing the
theory developed by C.C. Lin'' and ``forms a \textit{definite test}'' for it (Stromgren 1966b,
pp.3- 4; Stromgren 1967, pp.325, 329).

\bigskip

In response, Lin executed some ``preliminary explorations'' accounting for
the spiral component of galactic gravity, and found that ``even a small
spiral field [\ldots ] could be quite significant'' (LYS, p.734). He then
urged Chi Yuan to check ``whether there exist a pattern speed and a strength
of the spiral gravitational field (or a range for it) such that the stars
considered are found to have been formed in the gaseous arm as expected''
(Yuan 1969b, p.890). Experimenting with different choices of the parameters,
Yuan preferred a pattern speed of about 13.5 km/sec/kpc and a spiral field
strength of about 5{\%}, with which he calculated time-reversed motions of
25 stars from the Stromgren sample, their ages being `optimized', or
arbitrarily shifted within 15 percent in the desired sense. With these (and
several other) corrections Yuan succeeded in improving Stromgren's picture
and taking out of the interarm space all his stars that fell there \textit{but one}. LYS
proclaimed this result as offering an ``impressive agreement'' and ``also
extremely satisfactory comparison between theory and observation'' (LYS,
p.736). That was an overestimate, as was soon shown to the authors
(Contopoulos 1972; Kalnajs 1973) and as they conceded in turn.\footnote{ Contopoulos gave two reasons
why he did ``not consider this test as crucial''. First, he calculated the
uncertainty in the birthplaces, assuming an uncertainty in the ages of 10-15
percent, and found that that was large enough ``so that most of the stars
found by Yuan as born between the spiral arms may well have originated in a
spiral arm, \textit{without considering the attraction of the arm}''. Secondly,
he noticed, \textit{in any case} and any spiral galaxy the stars spend
on the average more time in the arms than between them. ``Therefore, finding
that the perturbed orbits give the places of origin in the spiral arms does
not provide a good test for the particular model chosen. [\ldots ] Similar
results were found by Kalnajs (private communication) after a more detailed
analysis''. (Contopoulos 1972, p.91)\par Indeed, Kalnajs (1973)
reproduced all of Yuan's calculations and determined their statistical
significance. He found that even when correcting star ages following Yuan in
a most advantageous manner, one to three stars from the latter's sample
should anyway be expected to be `bad' and not to leave the interarm
territory. Yuan had one such star\textit{ at least}. (Stromgren's initial sample included 26,
not 25 stars; the omitted 26$^{th}$ proved `bad', too.) Therefore,
Kalnajs concluded, Yuan's ``calculated birthplaces of the stars, while in
agreement with the expectations of the density wave theory, do not provide a
test for the presence of the spiral field'' (Kalnajs 1973, p.40).
``Perhaps C.C. thought this was a stringent test of the theory, but as I
discovered, the truth is quite the opposite: nothing really could have gone
wrong, and what little did go wrong was hushed up by the omission of the
errant star \#{26}.'' (\textit{Kalnajs}) \par ``I never responded to Kalnajs'
article, Yuan comments. The reason was he stressed the point, if I am not
mistaken, that we have not proven the density wave theory to be correct by star
formation study. We did not want to challenge that point. In fact, we agree
with it. We only demonstrated the consistency between the theory and
the observations (not only star migration but all other studies, e.g.,
streaming motions, vertex deviation, etc). I believe that his Observatory
article was written to respond to some of the strong claims of the density
wave theory made by C.C. in early days. [...]
\par My early contribution
to the density wave theory is to piece together all the relevant
observations to show the consistency of the theory. One aspect in agreement
is not enough, but the agreements with all observations are impressive. The
most significant early work for me was the doubly periodic solution of the
MHD density waves (Roberts and Yuan 1970; that paper was alphabetic order
in authorship; I made the crucial assumption and formulated the problem and
solved it in parallel to Roberts). That work was shortly confirmed by
Mathewson in observation of synchrotron radiation of M51. It produced a
strong support of the density wave theory. That MHD model is still the best
model for the Milky Way.'' (\textit{Yuan})} 

\bigskip
\bigskip

\section*{III. SHARPER FOCUS}
\addcontentsline{toc}{section}{III. Sharper focus}

\bigskip

{\footnotesize \begin{list}{}{\leftmargin4cm}
\item When a discovery is already done, it usually appears so evident that one
cannot but wonder why nobody hit upon it before.
\begin{flushright}
\textit{P.A.M. Dirac 1977}
\end{flushright}
\end{list}}

\bigskip

\subsection*{3.1 A feel of group velocity}
\addcontentsline{toc}{subsection}{3.1 \it A feel of group velocity}

\bigskip

{\footnotesize \begin{list}{}{\leftmargin4cm}
\item And yet, though it may be premature to speak of spiral waves as true modes
of oscillation, it seems entirely appropriate to ask how some postulated
spiral wave pattern in a galactic disk would \textit{evolve} with time.
\begin{flushright}
\textit{Toomre 1969, p.899}
\end{flushright}
\end{list}}

\bigskip

\noindent The WKBJ-style hot-disk dispersion relation admitted at least two different
treatments. Lin and Shu's rested on its `modal' form $\lambda (\nu )$.
Looking for a particular two-armed spiral wave, they let it rotate with some
angular speed $\Omega _{p} $, converted it to its pure-note frequency
$\omega = 2\Omega _{p} $, got it differentially `Doppler-shifted', $\omega
_{\ast}  = \omega - 2\Omega (r)$, found a ratio $\nu \equiv \omega _{\ast}
/ \kappa $ and, upon substituting it into $\lambda (\nu )$, obtained and
plotted the ready-to-serve interarm-spacing function $\lambda (r)$ and its
pitch-angle cousin $i(r) = \lambda (r) / \pi r$.

\bigskip

More in Lindblad's spirit, though equivalent, was the treatment stressing
that the dispersion relation specified the reduction of free oscillation
frequency $\kappa $ to some ${\left| {\nu}  \right|}\kappa $ due to
gravitational star coupling. This provided a deeper look at the so called
`dispersion orbits' -- ovals composed of many separate test stars at their
judiciously phased gyrations about a mean circumferential radius but devoid
of self-gravity. Such `orbits' precess at a rate $\Omega _{pr} (r) = \Omega
- \kappa / 2$, and if general rotation did ensure an approximate radial
independence of this combination, that alone would give practical prospects
of plaiting the happily co-revolving ovals into a common quasi-steady
two-armed pattern. Nature's choice proved slow variability of $\Omega -
\kappa / 2$, however, as if implying that there might be reason to try the
possibility of reducing $\Omega _{pr} (r)$ to a common value by allowing for
the as-yet-dormant star coupling. Indeed, that pointed at $\Omega _{pr} =
\Omega - {\left| {\nu}  \right|}\kappa / 2 = \Omega _{p}  = const$ with
its now Lin-Shu tuning formula $\nu (\lambda )$ for selecting spiral
geometries $\lambda (r)$, but to make this chance really work, it needed to
be demonstrated that the desired tuning of the precession rates could
actually be accomplished simultaneously, over a large radial span, and with
plausible interarm spacings.

\bigskip

Yet some restrictions were to be placed on these considerations. One was
that in a disk of stars the WKBJ waves could not abandon the territory
fenced by their related ILR and OLR. But Lin and Shu, who had rightly fixed
it in their 1964 patent, found this partial ban to be even a positive factor
as they let it favor the prevalence of two-armed spirals, on the simple
ground that only those might occupy the entire disk region between the best
separable $m$ = 2 Lindblad resonances. Still, for tentative disks reserving
local stability there happened to be another type of basic restriction.

\bigskip

Despite several early cautions (Kalnajs 1965, Julian 1967), Lin and Shu kept
on exploring their waves for the extra-helpful special case $Q = 1.0$ only.
This persistence seemed to annoy Toomre until late 1967 when he ``finally
ground out for [him]self what their dispersion relation would imply'' at $Q$>1
(\textit{Toomre}). He plotted $\nu (k)$ for different $Q$'s (Fig.8) and found that the case
$Q = 1.0$ was in a sense degenerate: it did let the WKBJ waves reach the
corotation circle from both sides, but just a minuscule addition to $Q$ was
enough to create their forbidden near-corotation zone that already for as
not so very much as $Q = 1.2$ paralyzed quite a sizable portion of the
disk.\footnote{ To be true, Shu was the first to discuss the $Q \ne 1$ disks
publicly, he did it in his thesis (Shu 1968) when attempting to ``finish
cataloging the nature of the dispersion relation for \textit{neutral} waves''
(S68, p.113). Because this nature ``changes somewhat when $Q \ne 1$'', he
``briefly summarize[d] [its] salient points'' (p.114) using a special plot
(Fig.9). For the $Q > 1$ that summary read: ``There is a
region about ${\left| {\nu}  \right|} = 0$ for which spatially oscillatory
waves cannot propagate. Toomre (private communication) has computed that for
values of $Q$ which are moderately greater than unity, the region of
inaccessible ${\left| {\nu}  \right|}$ can be quite substantial. [...]
Such an effect is not too serious since for pattern frequencies of the range
to be considered [...] the corresponding annular region where spatially
oscillatory waves cannot exist is small in comparison with the range where
they can exist. When $Q$ is greater than unity, the reflection, refraction, and
tunneling of propagating waves by and through such annular regions become a
serious problem for investigation'' (p.116). Shu, however, did not explain,
nor did he even hint, why he let this serious problem miss the threshold
case $Q $=1 where his proposed `Mode-A' was so welcome to cross corotation
smoothly, i.e. with no reflection, no refraction and no change in the sign
of group velocity.}

\begin{figure}[t]
\centerline{\epsfxsize=0.9\textwidth\epsfbox{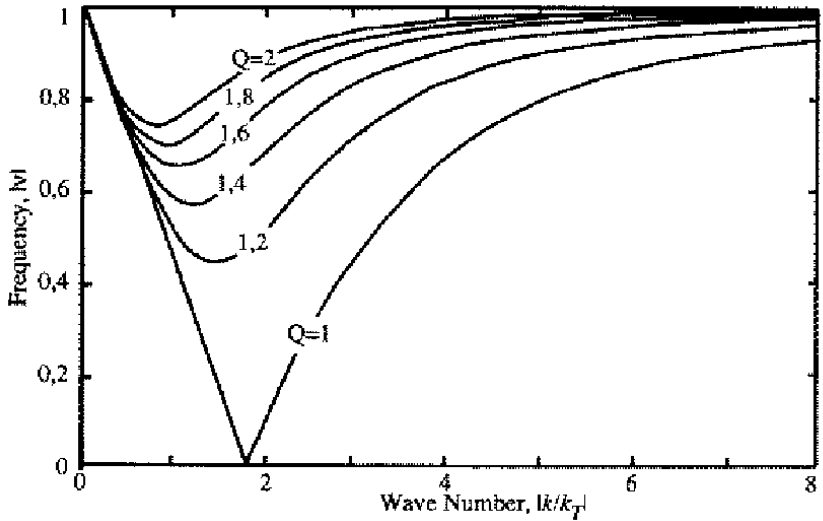}}
\caption{\footnotesize \textit{ The Lin-Shu-Kalnajs dispersion relation.}
Wavenumbers are in units of $k_{T} = 2\pi / \lambda _{T}$.
(The figure is reproduced from Toomre 1969)}
\end{figure}

\bigskip

Not even this, however, was the chief restriction to the envisaged
frequency-tuning success. ``Lin and Shu completely overlooked that in
repairing one serious defect they had actually created another: An
inevitable price for altering those speeds of precession in a
wavelength-dependent manner via the (very sensible) radial forces is a
\textit{group} velocity, likewise directed radially'' (Toomre 1977, p.449). Evaluation of
this `price' made the point of Toomre's work ``Group velocity of spiral
waves in galactic disks'' (Toomre 1969, hereinafter р69). In his preliminary
``Note on group velocity'' that in late 1966 was privately circulated at
MIT, Toomre had discussed the dispersive properties of the original cold
disk by Lin and Shu and called upon extending his discussion to their newer
and fairer hot model.\footnote{ As defined in the standard $d\omega / dk$
fashion, the `group velocity' of a rotating cold disk grows infinite as one
approaches the critical wavelength $\lambda _{T} $, below which the model
gets unstable.}  But Lin himself ``never took it very seriously'', and not
only because there were ``plenty of reasons not to brag about that old
note'' (\textit{Toomre}). More generally, at the time he fell into a muse over the role of
his introduced `reduction factor', when the group aspect might well appear
to him merely as an unnecessary tedious detour in the pursuit of his plain
ideas, so that he got no particular intention to `comb' the hairy and
transcendental Bessel functions and the like in his dispersion relation for
finding out some certain explicit function $\omega (k)$ just to take its
trite derivative.\footnote{ ``Besides, why in fact would anyone want to
differentiate the frequency $\omega $ only with respect to the radial
wavenumber $k$ instead of also the circumferential wavenumber $m$, since
`everyone knows' that a group velocity is a vector quantity, with components
in both directions? This question sounds pretty silly in retrospect, but
obstacles like that often seem a lot taller when they are first met.''
(\textit{Toomre})}

\begin{figure}[t]
\centerline{\epsfxsize=\textwidth\epsfbox{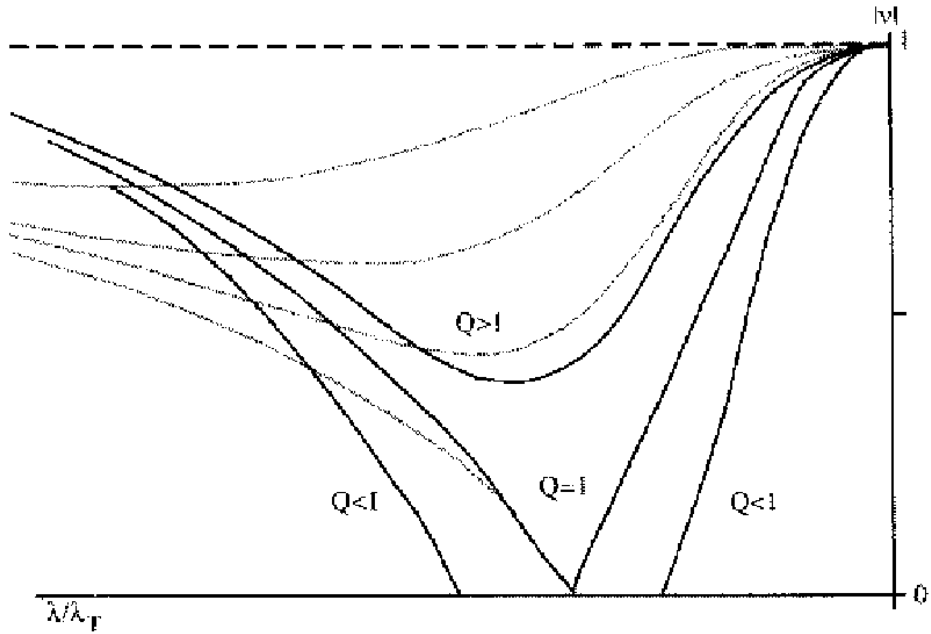}}
\caption{\footnotesize \textit{ The Lin-Shu-Kalnajs dispersion relation in
the `modal' form $\lambda (\nu).$} Shu's original curves (reproduced from
Shu 1968) are given against the exact (lighter) curves }
\end{figure}

\begin{quotation} {\footnotesize
\noindent \textit{Lin}: ``The general theory of group velocity is a well-developed and much taught
(e.g. in quantum mechanics) classical study valid for any dispersion
relationship connecting that wave number with the frequency. Different
people will feel differently whether it is even necessary to go beyond
taking the derivative and develop it anew for any each specific application.
I adopted the empirical approach. [\ldots ] That was directly related to the
calculation (or derivation) of the dispersion relationship. For this
calculation, Frank Shu did his share. The Lin {\&} Shu 1964 paper showed
that the crucial step is the calculation of the \textit{reduction factor}. [\ldots ] As I worked out
the dispersion relationship, I realized that the present problem is further
complicated by the presence of resonances. Thus the hope of success in the
calculation of modes depends on a very long-term effort (as it indeed turned
out to be the case). Thus our strategy not to pursue the dynamical approach
immediately turned out to be the right choice.''\footnote{ In 1968 Lin gave
a course at the Brandeis University Summer Institute, with the purpose
``to present the modern version of the \textit{density wave} theory as
developed over the past few years by myself and my collaborators
Frank H. Shu, Chi Yuan, and William W. Roberts'' (Lin {\&} Shu 1971, p.239).
Put on paper, that course appeared as Lin {\&} Shu 1971.
Speaking there of the spiral interest of prominent astronomers for many years,
the authors emphasized that ``until recently, however, there has always existed the
\textit{dilemma of differential rotation}'', and claimed that ``in this
article, [they] shall present an essentially stellar dynamical theory for the
\textit{persistence of the spiral pattern in the presence of differential
rotation}'' (Lin {\&} Shu 1971, p.239, 248). Then -- as originally in
Lin {\&} Shu 1964 -- they ``venture[d] to suggest that there are indeed
large-scale neutral (or nearly neutral) waves of spiral form for most of the
disk galaxies, and formulate[d] [their] ideas in the form of the [QSSS] hypothesis''.
They pointed out that accepting it and ``following the line of reasoning that
led to it'' one infers, among other things, that in the coordinate system rotating
with the pattern ``all phenomena are stationary'' and ``both the stream lines and
the magnetic field lines form closed nearly circular loops coinciding with each other''
(Lin {\&} Shu 1971, p.248). Then the authors gave a basic presentation of the WKBJ dispersion
relation and of the ensuing comparison with observations, and concluded
that their spiral theory ``needs extension in several directions to complete
the theoretical understanding of basic mechanisms and to develop its implications''.
The directions there envisaged were (1) the ``Thickness effect'', (2) the ``Complete
formulation of the theory'', (3) the problem of the ''Origin of galactic spirals''
that ``cannot be solved without using the complete formulation mentioned above'',
and (4) the ``Nonlinear theory'', becoming important as ``one looks beyond the
developments of the complete linear theory''. In topic (3) the authors enthused on the
promising ``preliminary indications'' and called for ``much work [that still]
remains to be done'', mentioning in a footnote that ``Toomre (private communication)
has also carried out studies involving the propagation of a group of waves and
their initiation by external agents'' (Lin {\&} Shu 1971, pp.287-289).} (\textit{Lin})

\bigskip

\noindent \textit{Shu}: ``I remember Lin telling me that he had group velocity well before the T69
paper on the subject; however, none of us then had any idea (a) what the
group velocity carried, and (b) why the concept would be relevant to
disturbances with a single value of the wave frequency.'' (\textit{Shu})

\bigskip

\noindent \textit{Toomre}: ``The reasons why one does or does not choose to attack some scientific
problem from a particular direction are rather `artistic' in nature, and
hard to make (or even hope to make) very sound and rational. [\ldots ]
Surely group velocity may be terribly `obvious' in retrospect to various
learned scholars, but I believe it did not seem so to Shu at the time he
struggled with his 2$^{nd}$-order WKBJ thesis at Harvard.'' (\textit{Toomre})
\par}
\end{quotation}

The main reason why Toomre took to the topic seriously in late 1967 only,
about two years after his work with Julian had been completed and the Lin
{\&} Shu 1966 paper published, was that in the global-mode context the
latter seemed to him to be no more than a trial exploration, and what he
judged vitally important and necessary to prove or disprove those authors'
`asymptotic' hopes was a full-fledged disk analysis.\footnote{ ``In essence,
I was there just echoing what Agris Kalnajs actually wrote in his
not-very-conclusive but nonetheless farsighted concluding chapter on
``Instabilities and Spiral Structure'' of his 1965 thesis. Yes, it seemed to
both \textit{him} and me at the time -- plus probably Hunter, Lynden-Bell, Lebovitz,
etc. -- that there was a lot of hard but very promising work to be done on
the `global' behavior of full-scale disks.'' (\textit{Toomre})\par}

\begin{quotation} {\footnotesize
\noindent ``In that climate of opinion it wasn't immediately evident to me, or to
anyone else, that one would learn much from the group velocity of those
short WKBJ waves that Lin and Shu (1966) were suddenly proposing. Of course
I was wrong there, but at least I can boast that by 1969 I myself had
repaired that oversight!'' (\textit{Toomre}) \par}
\end{quotation}

Toomre became the first to take real action on the evident understanding
that \textit{if} indeed a quasi-steady spiral mode can form in a galaxy disk, then it
does it via natural wave-packet evolution. And group velocity, he showed,
describes at least qualitatively how different kinds of information from the
packet are transmitted along the radius, being therefore directly related to
the maintenance of \textit{all} sorts of spiral patterns, even steady ones.

\bigskip
\bigskip

\subsection*{3.2 Group properties of tightly wrapped packets}
\addcontentsline{toc}{subsection}{3.2 \it Group properties of tightly wrapped packets}

\bigskip

{\footnotesize \begin{list}{}{\leftmargin4cm}
\item \ldots a shatteringly destructive article.
\begin{flushright}
\textit{Lynden-Bell {\&} Kalnajs 1972, p.1}
\end{flushright}
\end{list}}

\bigskip

\noindent Various properties of certain types of waves are described in a unified way,
regardless of the specific sort of the medium in which they propagate. Such,
for instance, are nearly plane -- weakly modulated -- waves ${} $ $\varphi \left(
{{\rm {\bf x}},t} \right) = A\left( {{\rm {\bf x}},t} \right)\cos {\left[
{S\left( {{\rm {\bf x}},t} \right)} \right]}$ whose amplitude $A\left( {{\rm
{\bf x}},t} \right)$ is much less dependent of its arguments than the phase
$S\left( {{\rm {\bf x}},t} \right)$. Their wave vector ${\rm {\bf k}} = -
\nabla S$ and frequency $\omega = \partial S / \partial t$
get connected through a link $\partial {\rm {\bf k}} / \partial t + \nabla
\omega = 0$ meaning conservation of the wave crests in number, their being
neither created nor annihilated. One more link is the common dispersion
relation $\omega \left( {{\rm {\bf x}},t} \right) = f{\left[ {{\rm {\bf
k}}\left( {{\rm {\bf x}},t} \right),\eta \left( {{\rm {\bf x}},t} \right)}
\right]}$ (with parametric $\eta $-dependence reflecting spatial
inhomogeneity). Together, these two connections form equations

\begin{equation}
{\frac{{\partial \omega}} {{\partial t}}} + {\frac{{\partial f}}{{\partial
{\rm {\bf k}}}}}\nabla \omega = - {\frac{{\partial f}}{{\partial \eta
}}}{\frac{{\partial \eta}} {{\partial t}}},
\quad
{\frac{{\partial {\rm {\bf k}}}}{{\partial t}}} + {\frac{{\partial
f}}{{\partial {\rm {\bf k}}}}}\nabla {\rm {\bf k}} = - {\frac{{\partial
f}}{{\partial \eta}} }\nabla \eta .
\end{equation}

\bigskip

\noindent Their characteristic curves coincide with the solution \textbf{x}($t)$ of the
equation

\begin{equation}
{\frac{{d{\rm {\bf x}}}}{{dt}}} = {\frac{{\partial f}}{{\partial {\rm {\bf
k}}}}};
\end{equation}

\bigskip

\noindent they are understood as \textit{rays}, in analogy with geometric optics.
Determined by the left-hand side of (2), vector \textbf{\textit{c}}$_{gr} $plays
as group velocity, with it information on $\omega $ and \textbf{\textit{k}} is
conveyed along the ray. If the medium is in general motion with a speed
\textbf{U}(\textbf{\textit{x}},$t)$, the waves are carried away. A co-moving
observer finds their frequency shifted, $\omega  = \omega _{\ast}  $+
\textbf{\textit{k}}\textbf{U} (asterisk marking the shifted quantity), and
the equations (1), (2) preserving their form. In particular, for the Lin-Shu
WKBJ waves they become

\begin{equation}
{\frac{{d\omega}} {{dt}}} = 0,
\quad
{\frac{{dk}}{{dt}}} = - \left( {{\frac{{\partial f_{\ast}} } {{\partial
r}}}} \right)_{k} ,
\quad
{\frac{{dr}}{{dt}}} = \left( {{\frac{{\partial f_{\ast}} } {{\partial k}}}}
\right)_{r} = c_{gr} ,
\end{equation}

\bigskip
\noindent right how Toomre wrote them having $k$ and $m$ as radial and azimuthal
wavenumbers, and $\omega = \omega _{\ast}  + m\Omega (r)$.

\bigskip

Equations (3) describe the radial transmission of the signals informing one
about invariable wave frequency and knowingly changing wavenumber. Toomre
computed them for an easy-to-use but realistic model with $Q = const$, $V =
r\Omega (r) = const$ where the rays $r(t)$ just repeat, in relabeled axes, the
form of the dispersion curve $\omega _{\ast}  (k)$. They are followed always
in the sense of growing $k$, because of which leading waves $(k < 0)$ can do
nothing but unwind while those trailing $(k > 0)$ wind up more and more.
Given by the local `slope' $dr / dt$, $c_{gr}$ changes its sign as the ray
reflects from the near-corotation barrier, and the Lin-Shu adopted
short-wave branch of the solutions has it negative inside corotation, or
directed inwards. A value $c_{gr} \cong $-10 km/sec that Toomre found for
the solar vicinity yields an estimate of few galactic years only for the
signal to travel from corotation to the ILR. To an already existing tightly
wound trailing pattern, an entire ray family may be compared in its part
after the near-corotation turning point, giving exhaustive knowledge on the
current and following dynamics of such a wave packet (Fig.10): its
information will simply be conveyed inward and gather all at the ILR where
the wave group velocity and pitch angle tend to zero.

\begin{figure}[t]
\centerline{\epsfxsize=0.9\textwidth\epsfbox{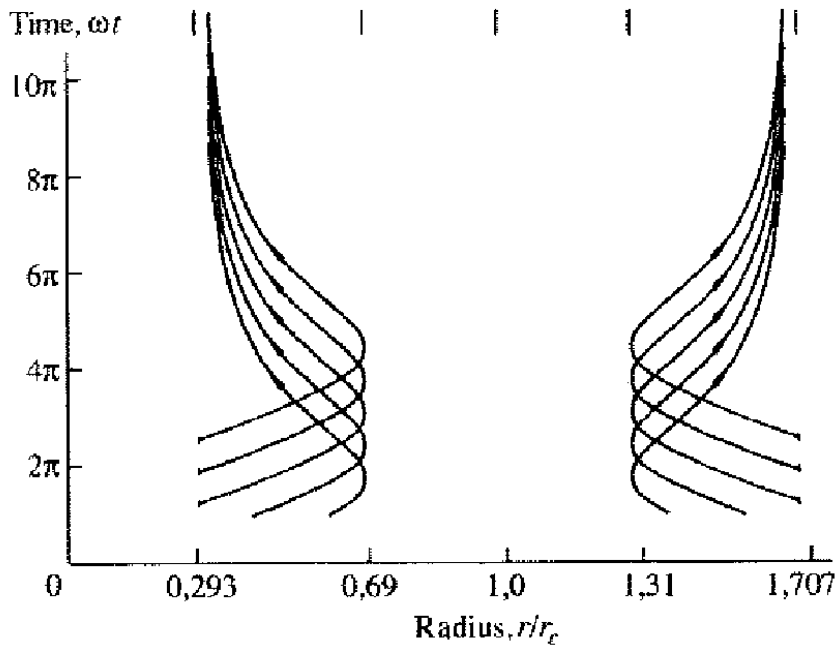}}
\caption{\footnotesize \textit {Some $m = 2$ characteristic curves (rays)
for a disk in which $Q = 1.2$} The $x$--axis is scaled
in corotation radius units. (The figure is reproduced from Toomre 1969)}
\end{figure}

\bigskip

To illustrate these `information' results, Toomre computed the wave-packet
evolution. There he relied on the program that had served him and Julian for
local needs of their Cartesian model (JT), because that model luckily
revealed an ability to mimic not only the corotation resonance $x = 0$ but
also the Lindblad resonances, since stars placed at and moving along lines
$x = \pm x_{L} = \pm \kappa/2Ak_{y} $ at the expense of shear were
ascertained to feel a $\cos (k_{y} y)$ wave at their natural frequency
$\kappa $.\footnote{ Of interest is the following record made by Toomre in
January 1968. ``The linearly shearing, constant surface density model of a
star disk that was used by Bill Julian and myself admittedly lacks i)
curvature, ii) boundaries, and iii) any gradients of unperturbed quantities
such as $c_{r}^{2} $ or $\kappa $. Nevertheless it can be used in the
following manner to illustrate to all desired numerical accuracy not only
C.C.'s dispersion relation for tightly wrapped spiral waves, but also the
related transient behavior and the transfer of energy. The point is that if
one were for some reason to choose any specific circumferential wave number
in the JT model, then as Agris correctly pointed out during Frank Shu's
thesis exam yesterday, our model, too, would have various Lindblad resonance
radii [\ldots and the region between them] will then correspond to what C.C.
and Frank call the `principal range'.'' (\textit{Toomre})} Toomre placed a short-term
emitter of such waves a little below $x = - x_{L} $ -- this imitated a bar
-- and purposely chose rather long-wave situations $\lambda _{y} / \lambda
_{T} \ge 4$ where the JT-exploited swing amplifier was all but shut off. His
computations showed that the `bar'-induced trailing-wave packet propagates
outwards; that its envelope drifts in approximate conformity with the
established by the ray methods characteristic curve; that indeed a larger
part of energy flow is reflected somewhere near corotation; and that the
packet drifts back to$x = - x_{L} $ where it eventually damps (Fig.11a).

\bigskip

The wave-packet evolution in the threshold $Q = 1.0$ disk was of particular
interest. While the Lin-Shu theory allowed the tightly wrapped waves to
reach and touch the corotation circle, it did not know if they could cross
it. And these waves showed they really could: the packet readily invaded all
the healthy tissue between the ILR and OLR, and even got amplified to a
degree, but then the inevitable group drift constricted it like a sausage at
$x = 0$, squeezed it out of that region and took the forming parts to their
LR destinations, as in the common case of $Q $ > 1 (Fig.11b).\footnote{ The
question on the preferable sense of spiral winding was not discussed
explicitly in T69. Toomre (as well as several others) held that its full
solution might be obtained only in the global-wave setting of the spiral
problem. At the same time, he was sure that several local findings already
gave a \textit{sufficient} understanding of the trailing-sense benefits. He meant, above all,
the delayed character of cooperative star wakes of non-axisymmetric forcing
from individual material clumps in a galaxy disk, and the group properties
of the Lin-Shu spirals. Indeed, since we do not observe them at the stages
of very loose winding and cross orientations, these stages either went
already (or were altogether absent) or they still shall have to go. In the
first case we have the trailing spirals whose old times are almost unknown
to us but whose long-lived future is unambiguously associated with states
pretty close to the today's one. In the second case, we would have the
leading spirals, huddling up to their ILR and extremely tightly wrapped in
order to avoid premature unwinding before too long. Besides, not to forget,
the waves of the short-length limit get excited with almost no concern of
self-gravity, so that only some `pressure'-force mechanism can generate
them. But what might be concretely any such \textit{`elastic'} mechanism localized in a narrow
circumcentral ILR region, and how would it manage to create a practically
circular wave running away ($c_{gr} $ > 0) from a gently sloping (inelastic)
`beach' of the ILR instead of rushing on it just like an ocean wave? Only
something akin to a Maxwell demon, Toomre guessed, could manufacture such
short leading waves. \par Yet he mentioned them once in T69 in the positive
sense. Speaking in a footnote of plausible variants for either one or both
$m = $2 Lindblad resonances to be absent from a galaxy disk, he remarked that ``in
such cases the given wave packet must in some sense be reflected either from
the outer edge of the disk or from its center'' and that ``in the process
the character will presumably change from trailing to leading, and the sign
of the group velocity should also reverse'' (T69, p.909). But, true, at that
time Toomre did not think seriously about any such conversion.}

\bigskip

Now what physically do the waves carry over the star disk and how do they do
it? This question was not trivial at the time. Only by the mid-1960s Whitham
had worked out a general variational principle for describing a wide class
of wave fields with dispersion. For weakly modulated packets it led to the
equation

\begin{equation}
{\frac{{\partial}} {{\partial t}}}{\frac{{E_{\ast}} } {{\omega _{\ast}} } }
+ \nabla
\left( {{\rm {\bf c}}_{gr} {\frac{{E_{\ast}} } {{\omega _{\ast}} } }}
\right) = 0,
\end{equation}
\begin{figure}
\centerline{\epsfxsize=0.8\textwidth\epsfbox{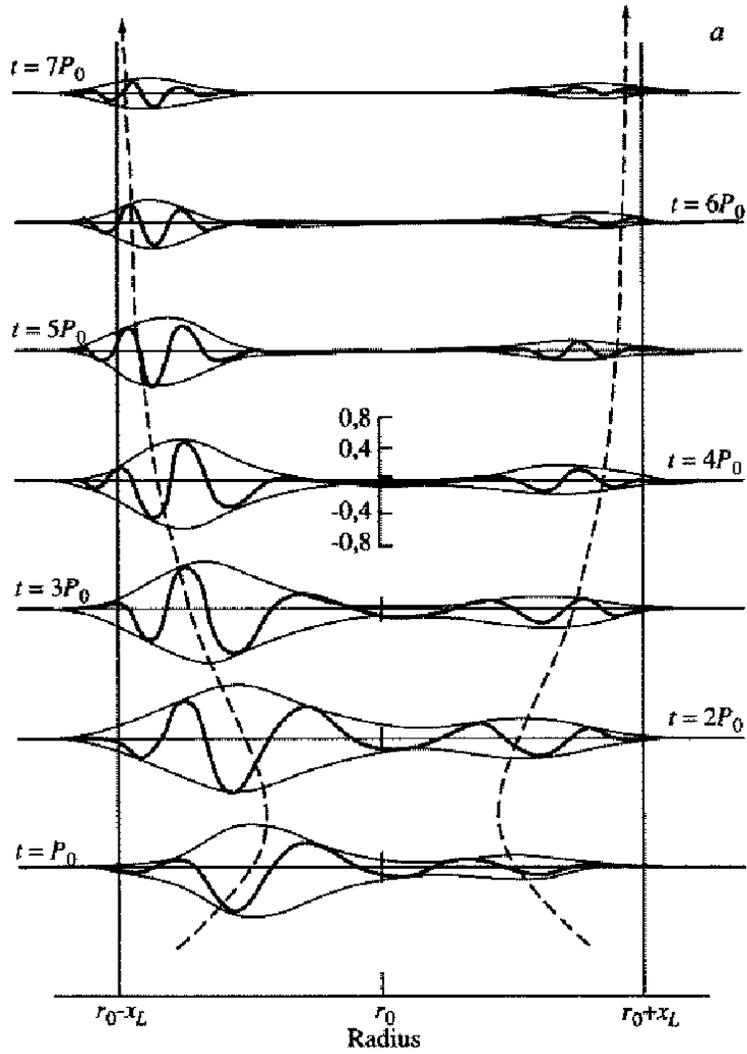}}
\caption{\footnotesize \textit{ A density wave evolving (a) in the $Q = 1.2$
and (b) $Q = 1.0$ local models.} $r_{0}$, $r_{0}$--$x_{L}$ and
$r_{0}+x_{L}$ are the corotation, ILR and OLR radii, $x_{L} = \lambda _{T}$,
$P_{0} = 2\pi / \kappa _{0}$. (The figure is reproduced from Toomre 1969)}
\end{figure}

\begin{figure}[t]
\centerline{\epsfxsize=0.8\textwidth\epsfbox{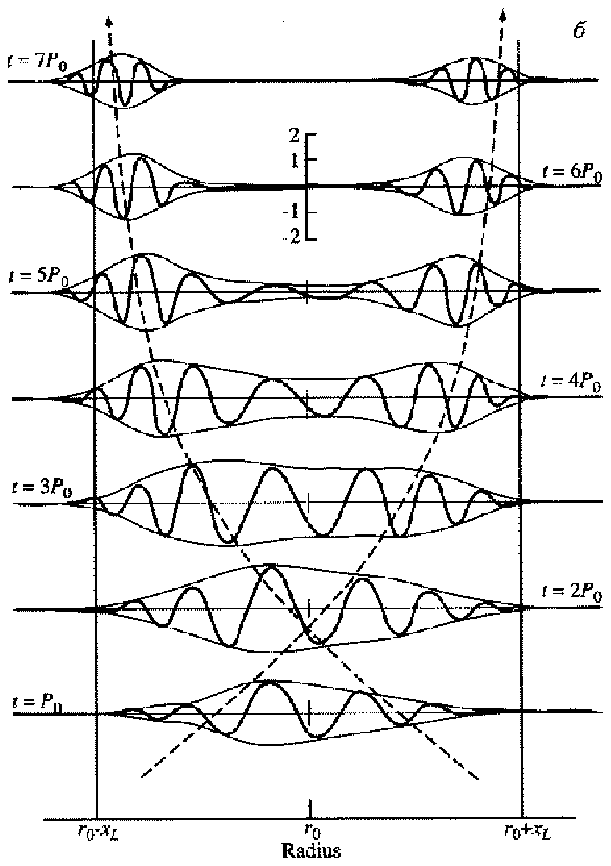}}
\centerline{Fig. 11 (b)}
\end{figure}

\bigskip

\noindent
expressing conservation of the wave-action density $E_{\ast}  / \omega
_{\ast}  $ and its along-ray transmission with group velocity ($E_{\ast}  $ being
the mean volume density of low-amplitude waves, and ${\rm {\bf c}}_{gr}
E_{\ast}  $ -- its flow). Toomre conceived that this should be applicable to
the Lin-Shu waves as well,\footnote{ ``I was glad enough to brag there that
I could also figure out that energy density itself, [... but] I was yet
prouder of noticing and pointing out that the main conserved density is not
even that energy as such, but instead the \textit{action} density [...] which Kalnajs
in turn soon told me had to be `the excess density of angular momentum
associated with the wave'. [...] There was nothing very original about
either accomplishment, though of course it could not have been entirely
obvious a priori that Whitham's Lagrangian reasoning would apply here as
well, with these collisionless stars rather than some more standard fluid.''
(\textit{Toomre})} and felt that the 2$^{nd}$-order WKBJ theory, which Shu had already been
developing to estimate the rates of change of wave amplitudes with radius,
should also yield an accurate radial derivative of $E_{\ast}  $. The
$dE_{\ast}  / dr$ that he first inferred from Shu's analyses differed in two
small but vital ways from that implied by equation (4). However Toomre
suspected that some errors had crept into Shu's work. In due course he
located them, and Shu soon concurred (Shu 1970b,c).\footnote{(\textit{Toomre}):
``Even in 1967 I was well aware that Frank Shu seemed to be progressing nicely
with his thesis, and was still claiming to confirm and to expand upon the `gradient
instability' which he and C.C. had announced rather cryptically in Lin {\&}
Shu 1966. In detail, I did not pay much attention until he had finished, but
then gave his analysis an exceptionally close going-over once he had been
awarded his PhD. [...] Amidst his immense and rather impressive
2$^{nd}$-order WKBJ calculation I eventually located two small algebraic
errors, once I had suspected because his inferred \textit{dE}/\textit{dr} did
not quite match what I had hoped for in what became eqn (34) of T69. Frank
soon agreed, and that was the end of those gradient instabilities!''
\par (\textit{Kalnajs}):
``As to the famous `gradient instabilities' I went as far as to type up
a short paper, dated July 29, 1968. I used my integral equation to show that
if you put corotation at the outer edge and made the same sort of tightly
wound approximations as Lin and Shu, then there could not be any instabilities.
But David Layzer thought that it would be far better if my first publication
on density waves made a positive contribution. So this effort remained in a
drawer. As it turned out, the 1969 Toomre paper made a positive contribution
to the field and at the same time debunked the `gradient instabilities'.''
\par Indeed, in a year Shu found it ``apparent that the growth (or decay) of
the wave amplitude arises because the disturbance is propagated radially in an
inhomogeneous medium and not because the disturbance is inherently overstable''
(Shu 1970c, p.110).}  After these small repairs, as Toomre remarked (T69, p.910),
Shu's work unwittingly closed the main logical gap of his own paper, and this
concluded his expose of the serious \textit{strategic} error by Lin and Shu --
their oversight of the group velocity.

\bigskip

\subsection*{3.3 Sources of spiral waves}
\addcontentsline{toc}{subsection}{3.3 \it Sources of spiral waves}

\bigskip

For one thing, Toomre's work р69 logically debunked the very principle of
the Lin-Shu theoretical construction, having shown that for all the
profundity of their core QSSS hypothesis their selfsustained-wave claim did
not immediately follow from their genuinely straightforward --
azimuthal-force-free -- `asymptotic' dispersion relation. Yet it also made
an important positive offer. The point was that wave-packet drifting and
damping still did not exclude the possibility itself of really long-lived
spirals, it only implied that ``if such patterns are to persist, the above
simply means that fresh waves (and wave energy) must somehow be created to
take the place of older waves that drift away and disappear''.

\begin{quotation} {\footnotesize
\noindent ``Where could such fresh and relatively open spiral waves conceivably
originate? The only three logical sources seem to be: (a) Such waves might
result from some relatively \textit{local} instability of the disk itself. (b) They may be
excited by tidal forces from \textit{outside}, such as from a companion or satellite galaxy.
(c) Or they might be a by-product of some truly large-scale (but not
necessarily spiral) distortion or instability involving an entire galaxy''
(T69, p.909). \par}
\end{quotation}

Thus Toomre simply formulated the evident -- but as yet unreleased --
necessity of establishing real mechanisms for maintaining spiral structure
in galaxies.

\bigskip

At the time, Toomre's particular interest lay in the tidal
mechanism.\footnote{ By the 1960s, the version of gravitational tides as
mainly causing the observed variety of `peculiar' forms of interacting
galaxies had been discredited, and what was brought to the forefront were
alternative considerations about magnetism, explosions, ejections, and just
as-yet-unknown `forces of repulsion', all kept at a level of hopes and
suspicions (the topic has been nicely reviewed in Toomre {\&} Toomre 1973).
During the decade, the tidal ideas were being gradually rehabilitated, but,
Toomre noticed (Toomre {\&} Toomre 1972, p.623], ``judging from the
reservations admitted by Zwicky (1963, 1967) despite his former use of words
like `countertide' and `tidal extensions' -- and especially from the
vehement doubts expressed by Vorontsov-Velyaminov (1962, 1964), Gold and
Hoyle (1959)], Burbidge, Burbidge and Hoyle (1963), Pikel'ner (1963, 1965),
Zasov (1967), and most recently by Arp (1966, 1969a,b, 1971) -- it has
usually seemed much less obvious that the basis of also such interactions
could be simply the old-fashioned gravity.'' Only in the early 1970s did the
tides find proper treatment (Tashpulatov 1969, 1970; Kozlov et al 1972);
that was a period of general recovery and renewal of interests to galaxy
dynamics.}  It arose after his and Hunter's work on bending oscillations
(modes) of finite-radius thin disks of a single gravitating material (Hunter
{\&} Toomre 1969). Among other things, that study hypothesized that the
bending of our Galaxy might be due to the vertical component of tidal force
during a possible close passage of the Large Magellanic Cloud (LMC). The
authors reckoned that their relatively slowly evolving $m = 1$ retrograde
responses were the only plausible candidates for the observed distortion.
This made them infer a very close passage at a perigalactic distance of
20-25 kpc and, to bring estimates into appreciable consistence, even claim
to favor a solar galactocentric distance $R_{O} \cong 8$ kpc instead of a
little too `ineffective' 10 kpc sanctioned at the time by the IAU. But
Hunter and Toomre ``were blissfully unaware'' of the work by Pfleiderer and
Siedentopf (1961; 1963) and ``also did not realize the undue sensitivity --
which those German authors had already implied -- of any such disk to the
horizontal components of the same tidal force during a \textit{direct} encounter of low
inclination'' (Toomre 1974, p.351).\footnote{ Pfleiderer reasoned that tidal
action should be much the strongest in the exposed and relatively slowly
rotating outer parts of the galactic disks where the mass density is small
and its self-gravity must be weak. He thus just neglected the latter and
treated the disk particle dynamics as the restricted three-body
problem, these three being the test particle and mass centers of the paired
galaxies. Such an over-idealization greatly simplified his computer work
(which still remained time-consuming since hundreds of trial encounters were
required for an understanding of the effects of various mass ratios, orbital
parameters and times and directions of viewing). \par ``These test-particle
calculations can, of course, be criticized for their total neglect of any
interactions between the various particles. However, this is not to say that
the self-gravity of these relatively low-density parts of the disk should
immediately have been of major importance, nor does it contradict our
qualitative picture about the evolution of the waves: For one thing, the
relatively sudden passage of the LMC should have induced roughly the same
initial velocities regardless of the subsequent \textit{disturbance} gravity forces from within
this system. And also, it seems that the principal effect of that latter
mutual attraction of the various disk particles should have been to enhance
the shearing discussed above, since in effect it would have reduced the
epicyclic frequency $\kappa $ and thus caused the wave speeds $\Omega -
\kappa / 2$ at the various radii to become more disparate.'' (T69, p.912)}
In a sense, Pfleiderer became Toomre's eye-opener,~and in the closing part
of T69 he already proposed that much of any spiral density wave in our
Galaxy might have evolved from vibrations set up during such a passage of
the LMC. Providing its orbital eccentricity $e \ge 0.5$, it would have spent
less than one galactic year traversing the nearest $90^{0}$ of
galactocentric longitude, and in the direct -- not retrograde -- case the
implied angular speed $\Omega _{s} $ would have roughly matched the speed of
advance, $\Omega - \kappa / 2$, of the slow $m = 2$ `dispersion orbit'.
``And that, coupled with the dominant $m = 2$ character of the tidal force
in the plane, means any direct close passage of the LMC should have been
very effective in exciting $m = 2$ oscillations of the Galaxy'' at a radius
where $\Omega _{s} = \Omega - \kappa / 2$. ``It also suggests that, even
with self-gravitation taken into account, the resulting `pattern speed'
should have been of the order of 10 km/sec/kpc'' (T69, p.911).

\begin{figure}[t]
\centerline{\epsfxsize=\textwidth\epsfbox{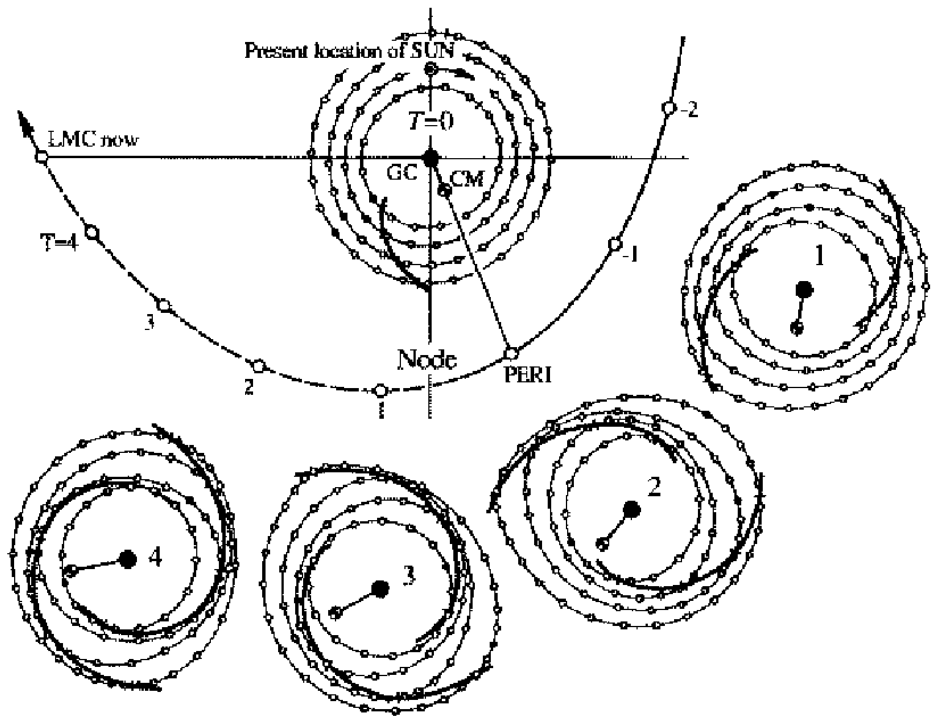}}
\caption{\footnotesize \textit{ A time history of the displacements of four
rings of noninteracting test particles provoked by a simulated
direct passage of the LMC.} The spiral curves connect points on
each ring which are at maximum distance from the Galaxy center.
Point `CM' marks the location of
the center of mass. Time in units of $10^{8}$ years is reckoned from the
perigalactic point. (The figure is reproduced from Toomre 1969)}
\end{figure}

\bigskip

Toomre (1969) supported his reasoning by computations of the perturber's
action on the Galaxy disk test particles (Fig.12). Then he made a separate
`progress report' at the Basel Symposium (Toomre 1970), but soon turned his
tidal interests to more spectacular and controversial forms, which resulted in
the famous dynamical study of `galactic bridges and tails' done jointly with
his brother (Toomre {\&} Toomre 1972).\footnote{ ``The hopes of Hunter and
myself that an unusually close passage of the LMC caused the well-known warp
of this Galaxy proved to be sadly in error.~I worked on that topic quite
intensely for another year or so, and even `predicted' a long tidal stream
to be torn loose from the LMC in turn [...] and probably inclined about 30
degrees to the plane of our Galaxy.~ I never published that, but it was well
enough known hereabouts that one day in early 1972 I got a sudden phone call
from Wannier or Wrixon at Bell Labs to ask whether a good chunk of what
turned out to be the Magellanic Stream which they had just then spotted --
about a year or two before Mathewson et al (1974) turned it into a big
business from Australia -- might possibly be the stream of gas that I had
asked about among several of our radio astronomers.~And I still remember
with pride that it took me just the few minutes during that phone call,
after learning this new Stream was located almost a right angles to this
Galaxy, to reply sadly that such an orientation or orbit could not help
Hunter and me at all, and that there apparently we had \textit{lost}! \par That Basel
example led me soon to enlist the help of my own brother Juri, who was then
affiliated with one NASA research institute in New York City that had much
better computers than any that I had access to ... and that in turn
eventually led to our joint paper Toomre {\&} Toomre 1972.~No, we did not
even come close to explaining the warp of our Galaxy ... but we did end up
explaining other nice things like NGC 4038/39 = `The Antennae', and pointing
out that the implied galaxy mergers probably explain why we have the
ellipticals. Quite a twist from where I began.'' (\textit{Toomre})}

\bigskip

\section*{IV. GATHERING IN BASEL}
\addcontentsline{toc}{section}{IV. Gathering in Basel}

\bigskip

\subsection*{4.1 Astronomers' applause}
\addcontentsline{toc}{subsection}{4.1 \it Astronomers' applause}

\bigskip

{\footnotesize \begin{list}{}{\leftmargin4cm}
\item Thus the credit goes to Lin who not only developed the theory of spiral
waves in much more detail, but also presented it in a relatively simple form
that made it acceptable to the rest of the astronomical world. The response
of the work of Lin and his associates has been an ever-growing wave research
in this area, that has produced many important new results.
\begin{flushright}
\textit{Contopoulos 1970a, p.303}
\end{flushright}
\end{list}}

\bigskip

\noindent The August 1969 Basel IAU Symposium ``The Spiral Structure of Our Galaxy''
was a significant event in the astronomical life, ``the first international
gathering ever of optical astronomers, of experts in galactic dynamics, and
of the world's greatest radio \textit{astrologers}''.\footnote{ An extract from \textit{Baseler Nachrichten} quoted in the
Symposium proceedings (Bok 1970).\par During the IAU General Assembly in
Prague, 1967, various theoretical and observational papers were presented at
a special meeting of Commission 33 on Spiral Structure, most notably
including ``The density wave theory of galactic spirals'' by Lin, ``Magnetic
approaches to spiral structure'' by Pikelner~and ``Self-gravitating spiral
models of the galaxy'' by Fujimoto. The participants' interest was obvious,
and Contopoulos proposed a special thematic symposium for 1969. It was
agreed to hold it ``in Basel, a center of galactic research in the center of
Europe'' (Becker {\&} Contopoulos 1970, p.vii).}  Bok, Contopoulos, Kerr and
Lin were the mainstay of its organization presided over by Woltjer who
deserved ``great credit for planning the symposium to reflect the current
status of our knowledge in this field, and for the selection of speakers''
(Lin 1971, p.35).

\bigskip

Opening the meeting, Oort conveyed his pleasure that Lindblad's spiral-wave
ideas had in recent years been ``further worked out by Lin, Shu and Yuan,
who showed among other things how such a density wave causing a spiral
pattern could be sustained by its own spiral gravitational field superposed
on the general axisymmetrical field of the galaxy'' (Oort 1970, p.1). This
gracious view the speaker supplemented with a prudent, if not veiledly
critical, comment.

\begin{quotation} {\footnotesize
\noindent ``The theory explains the maintenance but \textit{not} the \textit{origin} of spiral structure. I do
not think this is an important shortcoming, for it is easy to conceive of
processes which would \textit{start} a spiral structure. [\ldots ] A more serious problem
seems that of the \textit{long}-term permanence of the spiral waves. Can they continue to
run round during 50 revolutions without fatal damage to their regularity?
Looking at the irregularities in the actual spiral galaxies one wonders
whether the present spirals could \textit{continue} to exist for such a large number of
revolutions. [\ldots ]

\bigskip

Dr. Lin has sometime quoted me as having stated [\ldots ] that in so many
cases spiral arms can be followed more or less continuously through the
entire galaxy. I do not want to withdraw this statement, but I must point
out that it should be supplemented by two essential additions. First, that
in about half of the spirals the structure is either unclear, or there are
more than two arms. Second, that even in the half that can be classed among
the two-armed spirals there are invariably important additional features
\textit{between} the two principal arms, while the latter have often a number of secondary
branches coming off their outer rims.'' (Oort 1970, p.2) \par}
\end{quotation}

On its empirical side, the meeting revealed strong excitement and desire of
astronomers about establishing the Galaxy's spiral structure, at least in
general. Their demonstrations were a mixed collection, however. Even the
cutting-edge radio data instilled a scanty unanimity at best. Kerr (1970)
inferred the Perseus, Sagittarius, Norma-Scutum and Cygnus-Carina Arms as
spiral fundamentals, all of pitch angles $i = 5^{0} - 7^{0}$ (the latter
having our Sun at its inside, and the Orion Spur emanating from it), but
Weaver (1970) agreed on only the first two of them, and then with $i =
12^{0} - 14^{0}$. Was it to be wondered at the scatter of opinions of
`ordinary' optical reporters? Metzger (1970) found no definite spiral
pattern at all upon the distribution of HII regions. Courtes et al (1970)
re-interpreted data on radial velocities for about 6000 HII regions and
concluded an $i \cong 20^{0}$ four-armed spiral. Pavlovskaya and Sharov
(1970) gathered a 14-armed (!) spiral from their studies of surface
brightness distribution in the Milky Way plane. Vorontsov-Velyaminov (1970,
p.17) reminded that the largely discussed tightly wrapped two-armed spiral
proposal for our difficult Galaxy called for quite a number of full turns
inconsonant to the views of other galaxies, and he advised ``not to be in a
haste to construct a model of our Galaxy, but to search the real patterns
without bias''. Vaucouleurs (1970) interpreted the remarkable `3-kpc arm' as
a bar particularly oriented to the line of sight, for which he adduced even
more bar-favoring `statistical' arguments. Kerr (1970), to close this chain,
supported an oval distortion of the galactic plane as a plausible cause for
marked asymmetry of the observed rotation curve in the North and South
quadrants. At the same time, he expressed general concern over a large
uncertainty in the determining of galactic distances, which undermined the
cogency of any `large-scale' statements.

\bigskip

In this climate of empirical scatter and vagueness, Lin stated his ``bird's
eye view of theoretical developments'', as enlightened by his focal-point QSSS
hypothesis (Lin 1970). That view captured: ``ten general observational features
which one must consider in dealing with spiral features in galaxies''
(p.377); ``deep implications on the physical processes in the interstellar
medium, and in particular on the formation of new stars'' (p.379); ``a
better opportunity for the understanding of the physical processes,
including such microscopic behavior as the formation of molecules and dust
grains'' (p.389); ``a deep mystery of the 3-kpc arm'' that on fact might be
``a part of a reflected leading wave, of an evanescent type'' (p.383); the
necessity of recognition that the agreement with observations ``should \textit{not} be
perfect, since the galactic disk is perhaps not perfectly circular and the
actual structure may not be a pure mode in the theory'' (p.381); ''the
success of the theory'' as being expected ``to embolden us to apply the
theory to external galaxies'' (p.379). But Lin's special emphasis was there
and on ``one important theme to be kept in mind'' -- \textit{coexistence}.

\begin{quotation} {\footnotesize
\noindent ``The complicated spiral structure of the galaxies indicates the coexistence
of material arms and density waves, -- and indeed of the possible
coexistence of several wave patterns. When conflicting results appear to be
suggested by observations, the truth might indeed lie in the coexistence of
several patterns. Before taking this `easy way out', one should of course
try to examine each interpretation of the observational data as critically
as possible.

\bigskip There is also coexistence in the problem of origin of spiral structure. From
our experience with plasma physics, we learned that there are many types of
instabilities. Since a stellar system is basically a plasmoidal system,
various types of instability can also occur in the problem of the galactic
disk.'' (Lin 1970, p.379) \par}
\end{quotation}

Under the auspices of the coexistence theme and in grudging admiration for
Toomre's group-velocity work (as yet unpublished\footnote{ Contopoulos also
mentioned it in Basel, reviewing theoretical spiral developments (Contopoulos
1970a).}) that ``brings the problem even into sharper focus'' (Lin 1970, p.383),
Lin let the sheared waves \textit{coexist} in his grand-design view in order to
provide it with one of several possibilities of an instability mechanism. He
presumed that such waves naturally occur on the outskirts of a galaxy disk
where stars are sparse and the well-cooled gas is dominant; that this calls
up ``the Jeans instability of the galactic disk'' and thus a random occurence
of local condensations developing in the GLB fashion into trailing
spiral-shaped segments; and that each such `material arm' produces its own
effect, now in the JT manner, and ``eventually becomes a roughly
self-sustained entity, somewhat like the self-sustained density waves (Lin
{\&} Shu 1964, 1966; Lin 1966a) with inherent frequency $\nu = 0$
(corotating waves)'' (Lin 1970, p.384). But in fact such `entities',
established in local frames though, were on an intermediate scale at the
least, and far enough from their producing source they could indeed appear
``somewhat like the self-sustained density waves'', but with \textit{non-zero} frequency $\nu
$. Thus it was not entirely unreasonable to suspect that a really massive
outside perturber might be capable of bringing to life some grand galactic
spiral.\footnote{ Commenting on the fact that ``the M51 type spirals in
Vorontsov-Velyaminov's catalogue all have the companion galaxy or galaxies
on the arm'', Lynden-Bell well admitted that there ``the distortion gravity
field of the companion is very important and something on the lines of
[Toomre's] work with Julian ought to apply.`` (Lynden-Bell 1965b)} Lin,
however, was ``inclined to discount the roles of the satellite galaxies in
creating spiral patterns''. Believing instead that ``indeed, the corotation
of wave pattern and material objects in the outer parts of external galaxies
ha[d] been confirmed for M33, M51 and M81 by Shu and his associates (Shu et al
1971)'', he favored a picture where M51-type spiral patterns well originated
in remotely corotating local `self-sustained entities' and one of the arms
``would join naturally to the intergalactic bridge'' (Lin 1971, p.36-37).

\begin{quotation} {\footnotesize
\noindent ``Owing to resonance, the two-armed structure will prevail as the
disturbances propagate inwards as a group of waves, which extracts energy
from the basic rotation of the galaxy. [...] The reflection of the waves
from the central region then stabilizes the wave pattern into a
quasi-stationary form by transmitting the signal, via long-range forces,
back to the outer regions where the waves originated. Thus, there is
necessarily the coexistence of a very loose spiral structure and a tight
spiral structure. Population I objects stand out sharply in the tight
pattern while stars with large dispersive motion would primarily participate
in the very loose pattern.'' (Lin 1970, p.383)\footnote{ In one year or so
Lin's enthusiasm for this scenario will be tempered. He will concentrate on
mechanisms of spiral persistence, and direct his associates' efforts to
exploring the feedback cycles. Remote corotation will be as important there
as before, but now without reference to the GLB and JT ideas. The main point
will be the WKBJ-wave excursions to and from the center, and the associated
role of a central bar.} \par}
\end{quotation}

Lin's views of spiral structure, generously illustrated at the Basel symposium
in the coordinated presentations of his associates (Roberts 1970; Shu
1970a; Yuan 1970), evoked in quite a few of the astronomers a sort of delight
imparted so eloquently by Bok in his `Summary and Outlook'.\footnote{ Very
instructive is the view of the contemporary spiral progress by Goldreich who
after leaving the subject by mid-1960s ``remained an interested spectator to
the battles between Alar Toomre and C.C. Lin''. ``Although I generally
favored the arguments of the former -- he recalls -- the latter's campaign
was more successful.'' (\textit{Goldreich})}

\begin{quotation} {\footnotesize
\noindent ``Until half of a decade ago, most of us in this field were of the opinion
that the magnetic fields near the galactic plane [...] would probably
have proved sufficiently strong to hold the spiral arms together as magnetic
tubes. [...] Theory took a new turn about five years ago, when Lin and
Shu entered the field with the density wave theory. [...] The
magnificent work of the MIT group loosely headed by C.C. Lin has made the
pendulum of interpretation swing toward Bertil Lindblad's gravitational
approach, and this is wonderful indeed. [... It] is now in full bloom,
but we must not fool ourselves and think that all is done except the mopping
up. [...] There is controversy aplenty even within the MIT-Harvard
family and this is all to the good.

\bigskip

We are fortunate indeed that the theorists attended our Symposium in force.
[...] The observational astronomer is especially pleased to learn about
the interest our theoretical colleagues are showing in observations, and it
is a source of regret to the observers, optical and radio alike, that we
cannot agree as yet on the full outlines of spiral structure for our Galaxy.
Give us a few more years, and we shall be able to tell you all right!'' (Bok
1970, pp.457-462)\footnote{ Possibly, such a generous support of Lin's
initiative by several leading astronomers of the day partly reflected their
desire to see in him a direct follower of Lindblad, their previous
indifference to whose efforts might have evoked in them feelings of regret
and some guilt. ``I do not believe it -- Contopoulos comments on this guess.
-- In particular Bok wanted a simple theory to explain star formation and
migration. I remember that when I presented the work of Fujimoto in Prague
(1967) and wrote down only two formulae he told me: ``Very good George, but
too mathematical''. A few years later, Bok expressed his disappointment to
me, because the density wave theory had become rather complicated. I do not
think that Bok appreciated the more formal work of Lindblad.''
(\textit{Contopoulos})} \par}
\end{quotation}

\bigskip

\subsection*{4.2 Distinct cautions}
\addcontentsline{toc}{subsection}{4.2 \it Distinct cautions}

\bigskip

{\footnotesize \begin{list}{}{\leftmargin4cm}
\item Lin's programme for developing Lindblad's idea into a full theory has up to
now led to a theory of waves with neither a convincing dynamical purpose nor
a certain cause.
\begin{flushright}
\textit{Lynden-Bell {\&} Kalnajs 1972, p.25}
\end{flushright}

\item All things considered, only cumbersome `global' mode analyses and/or
numerical experiments seem to offer any real hope of completing the task of
providing the wave idea of Lindblad and Lin with the kind of firm
\textit{deductive} basis that one like to associate with problems of dynamics.
\begin{flushright}
\textit{Toomre 1977, p.452}
\end{flushright}
\end{list}}

\noindent Public acknowledgment of the Lin school was quite natural. Its initiative
greatly helped in re-orienting astronomers toward active recognition of and
observational tests for the gravitational nature and density-wave embodiment
of large-scale spiral structure. No sooner had Lin adopted the QSSS
hypothesis, he set himself the urgent and essential task of giving it
adequate empirical support. The thing demanded a practicable analytical
tool, and by 1966 he got it in a facile and handy asymptotic dispersion
relation. That it explained neither the origin of spiral structure, nor the
cause and mechanism for its tentatively long maintenance may well have
worried Lin, but in consort with his original plan he relegated these kinds
of topics to the future and rushed straight into empirical testing, having
added some heavy claims to his available basics as if adequately
backing the grand and quasi-steady spirals. Conveyed by him with the weight of
his authority, this played an important part in turning the tide of the
battle in his favor... and it affected the intuition, taste and attitude
of his audience toward more fundamental aspects of the spiral problem.

\bigskip

Nonetheless, there were presentations at the Basel meeting that alerted its
participants to the fact that true understanding of global spiral-making lay
far beyond the asymptotic theory they applauded and was bound to take quite
a while longer. One of the cautions came from Kalnajs (1970) in connection
with his long-term theme of coupled epicyclic oscillations of stars in a
thin disk.

\bigskip

Lindblad had introduced and studied the test-star-studded narrow rings --
`dispersion orbits'. Kalnajs (1965) in his thesis examined their
gravitational coupling, first in pairs\footnote{ Kalnajs considered a pair
of rings separated by a corotation region. He found that each of them is
corresponded by two basic oscillatory modes, one fast and the other slow,
and that even in axisymmetrically stable situations the different-type mode
coupling creates instability causing an outward angular momentum transfer.
Yet on this fact ``it would be premature to draw any conclusions about
spiral arms of galaxies'', he judged (Kalnajs 1965, p.81). ``By that time I
knew about the shearing sheet results. The two-ring example works even more
accurately in this setting. But of course one knows that the sheet is stable
and therefore the results inferred from two rings are not the same as that
from $2N$ rings -- of mass proportional to $1/N$ -- when one lets $N$ go to infinity.''
(\textit{Kalnajs})} and then in the whole, already in a continuous disk setting. There he
derived an integral equation for his disk's oscillatory dynamics,\footnote{To
reduce his complex integral equation, Kalnajs limited its frequency range
by specifying angular momentum radial distribution. He took Lindblad's
$\Omega - \kappa / 2 \cong const$ for the main part of a flat galaxy and the
Keplerian $\Omega \cong \kappa $ for its outer part, thus imitating (or
implying) an `edge' in his galactic system. As in the case of paired rings,
two modes, slow and fast, grew prevalent, the first one contributing much
more. This enabled Kalnajs to describe the modes separately and then account
for their coupling by perturbation theory methods. The kernel of the
slow-mode equation revealed no pole, it was symmetrical, and the mode stable
and devoid of trailing or leading signs. But the kernel of the fast-mode
equation had a pole at the OLR associated with the said `edge'. This changed
the qualitative situation: interacting with the OLR, the relatively slowly
growing perturbations supported the \textit{trailing} character of the fast mode and,
therefore, of the entire spiral wave.}$^{,}$\footnote{ (\textit{Contopoulos}): ``Kalnajs'
thesis~has a correct remark about trailing waves in a particular page. I
copied it and asked Toomre whether he could find there the preference
of trailing waves, but he couldn't. This convinced me that I should publish
my own results.'' \par (\textit{Toomre}): ``I have no such memory, but this is in no way to
dispute George's own recollection. [...] He always strove to be very
fair to Agris as a significant independent worker who had many good ideas
and sound mathematics. And so it is entirely plausible that he asked me
whether I thought that Agris -- then still lacking any true global-mode
results that his thesis had been struggling to develop -- had really
clinched that all realistic spirals must trail. Indeed, I remain pretty sure
that Agris by then had not done so \ldots but ask him yourself!''
\par (\textit{Kalnajs}): ``Unlike most people who would prefer a physical (or verbal)
explanation, George was keen to see the mathematics behind the
leading/trailing preference. Fortunately there is a simple enough
approximation of the galactic parameters in the vicinity of an OLR whose
contribution to the integral equation can be evaluated in closed form. The
result is eqn (117) of my thesis. In the subsequent three pages I explained
how that contribution to the kernel changes from one that in the absence of
a resonance does not favor leading over trailing, to one that prefers
trailing waves when a resonance is present. [...] Today I would use a
simpler example, perhaps the shearing sheet.'' } and in Basel he
demonstrated a variant of its numerical solution. That was a trailing
bar-spiral mode $m = 2$ with an e-fold growth in about two rotational
periods of the outer disk (Fig.13a). Kalnajs was pretty sure that his
analysis already resolved much of the global spiral-mode problem, and he
believed that at least qualitative confrontation with the evidence would
prove successful. In this respect he attached particular importance to the
fact that his analyzed gas-component reaction to the forming mode showed a
tightly wrapped two-armed spiral (Fig.13b). It was, however, far from
certain why his main unstable mode could not grow faster and how, even at
rather moderate growth rates, it would help one for very long. But anyway
that cast no doubt on Kalnajs' principal result -- the strong tendency of a
star disk to develop a temporary open two-armed spiral structure, which in
turn encourages bar-formation. Thus the Schmidt model of our Galaxy, which
was more or less favored in the 1960s by various investigators and which
Kalnajs now checked, was seriously unstable and unsatisfactory. If true,
this alone would soon overwhelm any `self-sustained modes' of Lin and Shu
peacefully revolving in a disk of stars.

\bigskip

Another caution came in Basel from the evidence provided by numerical experiments.
First computer simulations of the flat-galaxy dynamics as the $N-$body problem
had been performed in the late 1950s by P.O. Lindblad (1962). He then took
about 200 points only, because the early electronic computer was painfully
slow on direct calculation of paired interactions for appreciably higher
$N$'s. This stimulated new approaches to numerical experiments, and by 1968
Kevin Prendergast and Richard Miller worked out a more effective scheme
which, calculating forces in a limited number of cells, allowed rather quick
and accurate dynamical description of as much as 10$^{5} $particles or
so.\footnote{ Following Miller {\&} Prendergast (1968), particles `jumped'
between discrete-valued locations and velocities under discrete forces. The
fast finite Fourier-transform method was used for solving Poisson's equation
at each integration step.}  Inspired with the observation that ``because the
program is new, new results are coming rapidly'' (Prendergast {\&} Miller
1968, p.705), they and William Quirk prepared for Basel a motion picture of
``the very interesting physical implications'' of their experiments (Miller
et al 1970a,b). Spectacular spiral patterns were found to ``nearly always
develop [already] in the early stages'' of their model disks, yet -- they
argued -- these ``cannot be valid $N-$body analogues of the spiral patterns of
actual galaxies'' as most evidently reflecting violent reorganization of the
artificially arranged initial state. More importantly, it was ascertained
that ``machine calculations typically produce `hot' systems that are largely
pressure-supported'' (Miller et al 1970b, p.903-4), in contrast to the
observed thin disks in galaxies.

\begin{figure}
\centerline{\epsfysize=0.8\textheight\epsfbox{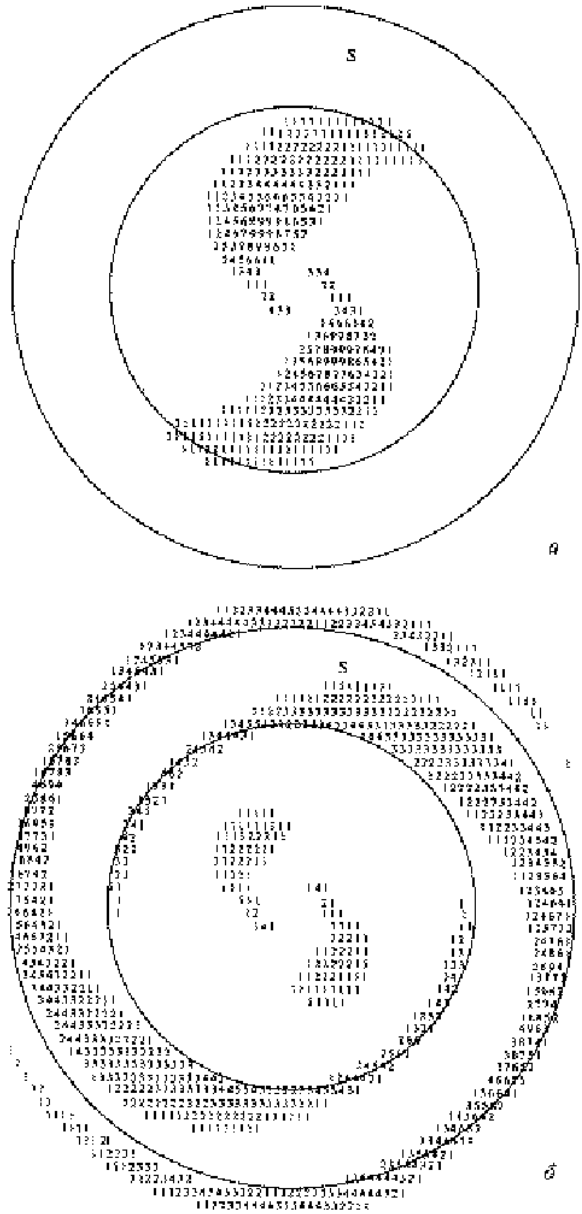}}
\caption{\footnotesize \textit{ Kalnajs' growing bar mode: excess densities
(a) in the stellar disk, (b) in its gas layer.} Large and small circles mark
the outer Lindblad and corotation resonances, point $S$ -- Sun's position.
(The figure is reproduced from Kalnajs 1970)}
\end{figure}

\bigskip

The above experimenters found a simple but interesting way out -- a `manual'
cooling ``by appropriately modifying the systems already in the computer''
(Miller et al 1970b, p.904). By integration steps (cycles) they cooled some
10{\%} of their particles, preserving their orbital momentum to imitate
their inelastic mutual collisions and make them dynamically akin to
interstellar gas clouds. As before, the remaining 90{\%} heated up to the
circular-speed-comparable velocity dispersions, but with this a bar was
formed and also a trailing pattern of moderate winding that, although
chaotic and flexible as it might appear as a whole, contained a bar-bound $m
= 2$ spiral wave component (Fig.14). Slowly revolving in the sense of
general flow, the bar and the gradually tightening spiral faded from the
sight in about three disk rotations.

\bigskip

Frank Hohl and Roger Hockney worked out a more accurate computational scheme
in the late 1960s (Hockney {\&} Hohl 1969). Unlike Prendergast and
coworkers, Hohl's interest lay in `pure' dynamics of collisionless
models.\footnote{ To avoid computational artifacts, Hohl had carefully
examined properties of his numerical schemes and showed that his $N-$body models
were indeed collisionless (Hohl 1973) and their behavior was independent of
the particle number, cell size and integration time step (Hohl 1970b).} In
Basel he experimentally confirmed the fact of fast -- for a period of one
revolution -- small-scale fragmentation of cold star disks and its
prevention by a massive (no less than four disk masses) spherical halo (Hohl
1970a). The hot disks were checked separately (Hohl 1971) to get stable in
Toomre's axisymmetric sense at the initial `temperature' $Q = 1$, but then
they still remained unstable against relatively slowly growing large-scale
disturbances that caused the system to assume a very pronounced bar-shaped
structure after two rotations (Fig.15);\footnote{ In two more rotations, a
nearly axisymmetric distribution of stars around a massive central oval
resulted, revolving about half as fast as the initial disk.}  $^{} $in major
features it confirmed the growing bar-spiral mode picture that Kalnajs
(1970) had obtained via his integral equation. The total lack of spiral
shapes of respectable duration in this and every other purely
stellar-dynamical experiment conducted with sizable fractions of `mobile'
mass was a result that almost spoke for itself.\footnote{ ``It is
conceivable, of course, that some milder instabilities which might
themselves have led to more enduring spirals, were thwarted in these
experiments by a kind of overheating from the fierce initial behavior. This
seems unlikely, however, because of Hohl's extra tests with that artificial
cooling (Hohl 1971).'' (Toomre 1977, p.468)} ~Indeed, just like the modal
work of Kalnajs and the N-body results of Prendergast et alia, it cautioned
everyone at Basel that this strong tendency toward bar-making very much
needed to be understood and tamed lest it overwhelm the QSSS hopes of Lin
and all his admirers.

\begin{figure}[t]
\centerline{\epsfxsize=0.8\textwidth\epsfbox{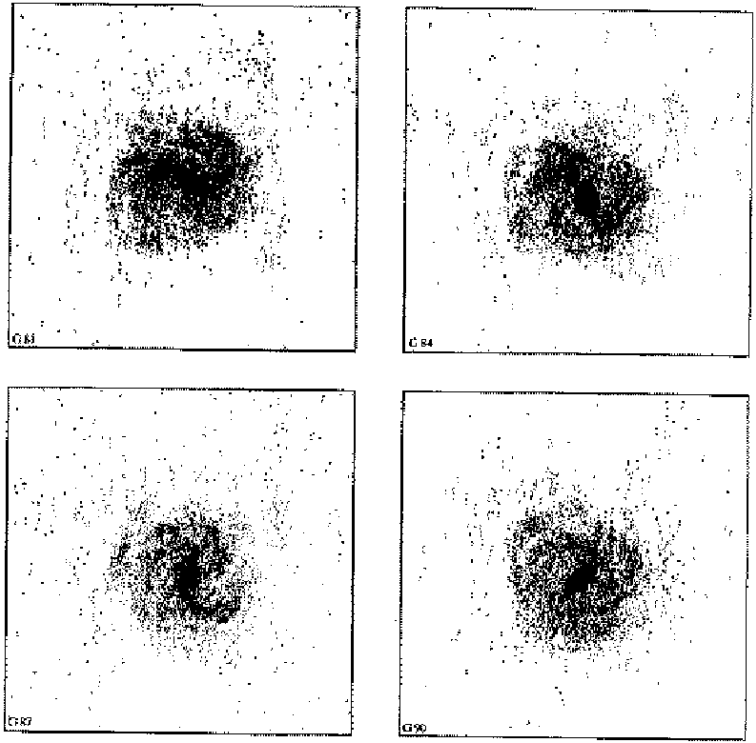}}
\caption{\footnotesize \textit{ The formation and evolution of the bar-spiral structure
in a partially cooled gravitating disk.}
(The frames are reproduced from Miller et al 1970)}
\end{figure}

\begin{figure}
\centerline{\epsfxsize=0.9\textwidth\epsfbox{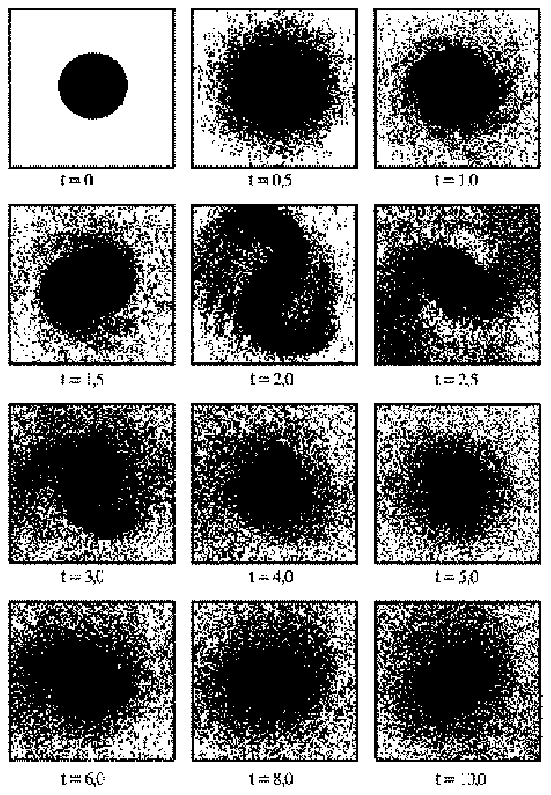}}
\caption{\footnotesize \textit{ The evolution of a stellar disk from an initially balanced
state of uniform rotation and marginal stability $Q = 1$.}
Time is in initial rotation period units.
(The figure is reproduced from Hohl 1971)}
\end{figure}

\bigskip
\bigskip

\section*{Afterword}
\addcontentsline{toc}{section}{Afterword}

\bigskip

By the beginning of the 1970s the spiral subject was in considerable
disarray. The still popular QSSS hypothesis of Lin and Shu, along with their
illustrative semi-empirical theory, was confronted with serious
difficulties. Lin and his associates were put clearly on the defensive over
their tightly wrapped (quasi)-steady modes on two principal fronts: from the
radial propagation at the group velocity that would tend to wind them almost
at the material rate, and from the tendencies of galaxy disks toward a
strong global instability that appeared likely to overwhelm them. Of course,
one might claim that all such threats were just imaginary and temporary, and
only of academic interest, on the ground that nature itself had overcome
them (as say for the case of the bar-making instability of stellar disks,
the rescue from which was actively sought in the 1970s in a massive inert
halo that in fact was not needed). One might also be confident that the QSSS
hypothesis must be correct, as illuminated by the everlasting truth of
Hubble's classification of the galactic morphologies. One might even take
pride in the historical fact that an interesting and very promising concept
developed, although not connected to the wave steadiness, on spiral shocks
in interstellar gas and their induced star formation. But such a heuristic
approach did not stimulate very strong progress in understanding dynamical
principles of the spiral phenomenon; moreover, it often misled, and a rich
irony was already that the supposed QSSS favorites M51 and M81 (Lin held
originally that a large majority of the galaxies -- 70{\%} -- ``are normal
spirals like the whirlpool'' (Lin 1966a, p.877)) turned out most probably
not to be quasi-steady at all. A further irony was the continuing failure of
Lin and Shu to account the trailing character of their `modes', while that
was already grasped by their direct `deductive' opponents. But the greatest
irony lay in the fact that the concept later known as swing amplification,
worked out by the mid-1960s, was originally denigrated by Lin's camp as
relating exclusively to `material arms', whereas it turned out in the end to
be of vital importance to this entire spiral enterprise including the
variants of chaotic ragged patterns, tidal transient grand designs and
growing or quasi-steady modes.

\bigskip

The 1970s that came promised many interesting events in the spiral arena,
because -- here we repeat what we said in the beginning of the paper and
with it close our narrative -- by that time it had become very clear to
everyone that much hard work still remained to explain even the persistence,
much less the dynamical origins, of the variety of spirals that we observe.

\bigskip
\bigskip
\subsection*{Acknowledgements}

\bigskip

The present Papers I and II were made possible
thanks entirely to the generous and responsive participants of the events
described. V. Antonov, G. Contopoulos, P. Goldreich, C. Hunter, W. Julian,
A. Kalnajs, C.C. Lin, P.O. Lindblad, D. Lynden-Bell, F. Shu, A. Toomre and
C. Yuan provided me with important materials, memories, opinions, and also
gave me, kindly and patiently, answers and comments on my endless queries.
Various help also came from L. Athanassoula, G. Bertin, H. Eisenberg, A.
Fridman, I. Genkin, G. Idlis, V. Korchagin, I. Korshunova, G. Kulikov, G.
Kurtik, M. Maksumov, Yu. Mishurov, D. Muhamedshin, M. Orlov, L. Osipkov, V.
Polyachenko, A. Rastorguev, M. Roberts, E. Ruskol, J. Sellwood, K. Semenkov,
V. Sheremet, F. Tsitsin. My special thanks go to T. Agekian, V. Gorbatsky
and V. Orlov for their kind invitations for me to speak at their seminars at
the St. Petersburg State University, and also to E. Kolotilov and V.
Komissarov for their hospitality during the period I was completing the
manuscript at the Crimean station of the Sternberg State Astronomical
Institute in the summer of 2003.

\newpage

\section*{References}
\addcontentsline{toc}{section}{References}

\bigskip

{\footnotesize

{\parindent=0pt

(\textit{Antonov}) = V.A. Antonov. Private communications, 2003.

(\textit{Contopoulos}) = G. Contopoulos. Private communications, 2000-03.

(\textit{Goldreich}) = P. Goldreich. Private communications, 2002.

(\textit{Julian}) = W.H. Julian, Private communications, 2002.

(\textit{Kalnajs}) = A.J. Kalnajs, Private communications, 2001-03.

(\textit{Lin}) = C.C. Lin. Private communications, 2000-01.

(\textit{Lynden-Bell}) = D. Lynden-Bell. Private communications, 2000-03.

(\textit{Shu}) = F.H. Shu. Private communications, 2001.

(\textit{Toomre}) = A. Toomre. Private communications, 2000-03.

(\textit{Yuan}) = C. Yuan. Private communications, 2001.

}

\bigskip

{\leftskip=0.25in\parindent=-0.25in

Princeton 1961 = The Distribution and Motion of Interstellar Matter in
Galaxies (Proc. Conf. Inst. Adv. Study, Princeton NJ, 1961). L. Woltjer, ed.
W.A. Benjamin, NY, 1962.

Noordwijk 1966 = Radio Astronomy and the Galactic System (Proc. IAU Symp. No
31, Noordwijk 1966). H. van Woerden, ed. London and NY, Academic Press,
1967.

Basel 1969 = The Spiral Structure of Our Galaxy (Proc. IAU Symp. No 38,
Basel 1969). W. Becker and G. Contopoulos, eds. Dordrecht-Holland, D.Reidel
Publ., 1970.

\bigskip
\bigskip

Antonov, V.A. 1960. \textit{Remarks on the problem of stability in stellar dynamics}. Astron. Zh. \textbf{37}, 918-926.

Antonov. V.A. 2003. Private communication.

Arp, H. 1965. \textit{On the origin of arms of spiral galaxies}. Sky and Telescope \textbf{38}, 385.

Arp, H. 1966. \textit{Atlas of Peculiar Galaxies}. Pasadena, California Institute of Technology.

Arp, H. 1969. \textit{Companion galaxies on the ends of spiral arms}. Astron. Astrophys. \textbf{3}, 418-435.

Arp, H. 1971. \textit{Observational paradoxes in extragalactic astronomy}. Science \textbf{174}, 1189-1200.

Athanassoula, E. 1984. \textit{The spiral structure of galaxies}. Phys. Rep. \textbf{114}, 319-403.

Becker, W., Contopoulos, G. 1970. \textit{Introduction}. Basel 1969, vii-viii.

Bertin, G. 1980. \textit{On the density wave theory for normal spiral structure}. Phys. Rep. \textbf{61}, 1-60.

Bertin, G., Lin, C.C. 1996. \textit{Spiral Structure in Galaxies. A Density wave Theory}. Cambridge MA, MIT Press.

Bok, B.J. 1970. \textit{Summary and outlook}. Basel 1969, 457-473.

Burbidge, E.M., Burbidge, G.R., Hoyle, F. 1963. \textit{Condensations in the intergalactic medium}. ApJ \textbf{138}, 873-888.

Burton, W.B. 1966. \textit{Preliminary discussion of 21-cm observations of the Sagittarius arm and the systematic motion of the gas near its edge.} Bull. Astron. Inst. Netherl. \textbf{18}, 247-255.

Chandrasekhar, S. 1942. \textit{Principles of stellar dynamics}. Chicago IL, Univ. of Chicago Press.

Contopoulos, G. 1970a. \textit{Gravitational theories of spiral structure}. Basel 1969, 303-316.

Contopoulos, G. 1970b. \textit{Preference of trailing spiral waves}. ApJ \textbf{163}, 181-193.

Contopoulos, G. 1972. \textit{The dynamics of spiral structure. Lecture notes}. Astronomy Program and Center for Theor. Phys. Univ.
of Maryland.

Contopoulos, G., Stromgren, B. 1965. \textit{Tables of Plane Galactic Orbits}, NY, Inst. for Space Studies.

Courtes, Y.P., Georgelin, Y.M., Monnet, G. 1970. \textit{A new interpretation of the Galactic structure from HII regions}. Basel 1969, 209-212.

Crawford, D.L., Stromgren, B. 1966. \textit{Comparison of the Hyades, Coma and Pleiades clusters based on photoelectric u, b, v, y and H photometry}. Vistas in Astron. \textbf{8}. A. Beer,
ed., Pergamon Press, 149-157.

Dirac, P.A.M. 1977. \textit{Recollections of an Exciting Era}. In: History of Twentieth Century Physics (Proc. Int.
School Physics ``Enrico Fermi'', Course LVII). NY - London, Academic Press,
109-146.

Drury, L.O'C. 1980. \textit{On normal modes of gas sheets and discs}. MNRAS \textbf{193}, 337-343.

Eggen, O.J., Lynden-Bell, D., Sandage, A.R. 1962. \textit{Evidence from the motions of old stars that the Galaxy collapsed}. ApJ \textbf{136},
748-766.

Fujimoto, M. 1968. \textit{Gas flow through a model spiral arm}. In: Non-stationary phenomena in galaxies (IAU Symp No
29, Burakan, May 1966). Erevan, Armenian SSR Acad. Sci. Publ., 453-463.

Gold, T., Hoyle, F. 1959. \textit{Cosmic rays and radio waves as manifestations of a hot universe}. In: Paris Symposium on Radio Astronomy, IAU
Symposium No. 9 and URSI Symposium No. 1, held 30 July - 6 August, 1958.
R.N.Bracewell, ed., Stanford CA, Stanford Univ. Press, 583-588.

Goldreich, P., Lynden-Bell, D. 1965a. \textit{Gravitational stability of uniformly rotating disks}. MNRAS \textbf{130}, 97-124.

Goldreich, P., Lynden-Bell, D. 1965b. \textit{Spiral arms as sheared gravitational instabilities}. MNRAS \textbf{130}, 125-158.

Goldreich, P., Julian, W.H. 1969. \textit{Pulsar electrodynamics}. ApJ \textbf{157}, 869-880.

Goldreich, P., Tremaine, S. 1978. \textit{The excitation and evolution of density waves}. ApJ \textbf{222}, 850-858.

Goldreich, P., Tremaine, S. 1979. \textit{The excitation of density waves at the Lindblad and corotation resonances by an external potential}. ApJ \textbf{233}, 857-871.

Goldreich, P., Tremaine, S. 1980. \textit{Disk-satellite interactions}. ApJ \textbf{241}, 425-441.

Hockney, R.W., Hohl, F. 1969. \textit{Effects of velocity dispersion on the evolution of a disk of stars}. AJ \textbf{74}, 1102-1104.

Hoerner, S. von, 1962. Princeton 1961, 107.

Hohl, F. 1970a. \textit{Computer models of spiral structure}. Basel 1969, 368-372.

Hohl, F. 1970b. \textit{Dynamical evolution of disk galaxies}. NASA Technical Report, R-343.

Hohl, F. 1971. \textit{Numerical experiments with a disk of stars}. ApJ \textbf{168}, 343-359.

Hohl, F. 1973. \textit{Relaxation time in disk galaxy simulations}. ApJ \textbf{184}, 353-360.

Hunter, C. 1963. \textit{The structure and stability of self-gravitating disks}. MNRAS \textbf{126}, 299-315.

Hunter, C. 1972. \textit{Self-gravitating gaseous disks}. Ann. Rev. Fluid Mech. \textbf{4}, 219-242.

Hunter, C., Toomre, A. 1969. \textit{Dynamics of the bending of the Galaxy}. ApJ \textbf{155}, 747-776.

Julian, W.H. 1965. \textit{On the Enhancement of Radial Velocities of stars in Disk-Like Galaxies}. PhD Thesis, MIT (August 1965).

Julian, W.H. 1967. \textit{On the effect of interstellar material on stellar non-circular velocities in disk galaxies}. ApJ \textbf{148}, 175-184.

Julian, W.H., Toomre, A. 1966. \textit{Non-axisymmetric responses of differentially rotating disks of stars}. ApJ \textbf{146}, 810-830 (JT).

Kalnajs, A.J. 1963. \textit{Spiral Structure in Galaxies. Outline of a thesis}. Harvard Univ., Cambridge, MA (October 1963).

Kalnajs, A.J. 1965. \textit{The Stability of Highly Flattened Galaxies}. PhD Thesis, Dept of Astron., Harvard Univ., Cambridge,
MA (April 1965).

Kalnajs, A.J. 1970. \textit{Small amplitude density waves on a flat galaxy}. Basel 1969, 318-322.

Kalnajs, A.J. 1971. \textit{Dynamics of flat galaxies. I}. ApJ \textbf{166}, 275-294.

Kalnajs, A.J. 1972. \textit{The damping of the galactic density waves by their induced shocks}. Astrophys. Lett. \textbf{11}, 41-43.

Kalnajs, A.J. 1973. \textit{Star migration studies have not yet revealed the presence of a spiral density wave}. Observatory \textbf{93}, 39-42.

Kerr, F.J. 1962. \textit{Galactic velocity models and the interpretation of 21-cm surveys}. MNRAS \textbf{123}, 327-345.

Kerr, F.J. 1970. \textit{Spiral structure of neutral hydrogen in our Galaxy}. Basel 1969, 95-106.

Kozlov, N.N., Sunyaev, R.A., Eneev, T.M. 1972. \textit{Tidal interaction of galaxies}. Dokl. Akad. Nauk SSSR
\textbf{17}, 413.

Ledoux, P. 1951. \textit{Sur la stabilite gravitationelle d'une nebuleuse isotherme}. Ann. d'Astrophys. \textbf{14}, 438-447.

Lin, C.C. 1966a. \textit{On the mathematical theory of a galaxy of stars} (June, 1965 Courant Symp., Courant Inst. Math. Sci., NY).
J. SIAM Appl. Math. \textbf{14}, 876-920.

Lin, C.C. 1966b. In \textit{Outline of talks presented at the Columbia November 4, 1966 Meeting, from notes taken by F. Shu}.

Lin, C.C. 1967a. \textit{Stellar dynamical theory of normal spirals} (1965 Summer school, Cornell Univ. Ithaca, NY). Lect. Appl.
Math. \textbf{9}, 66-97.

Lin, C.C. 1967b. \textit{The dynamics of disk-shaped galaxies}. Ann. Rev. Astron. Astrophys. \textbf{5}, 453-464.

Lin, C.C. 1968. \textit{Spiral structure in galaxies}. In: Galaxies and the Universe (October 1966 Vetlesen
Symp., Columbia Univ.). L. Woltjer, ed., Columbia Univ. Press, 33-51.

Lin, C.C. 1970. \textit{Interpretation of large-scale spiral structure}. Basel 1969, 377-90.

Lin, C.C. 1971. Sky and Telescope \textbf{42}, 35-37.

Lin, C.C. 1975. \textit{Theory of spiral structure}. In: Structure and Evolution of Galaxies. G. Setti, ed., D.
Reidel Publ. Comp., 119-142. The same in: Theoretical and applied mechanics (Proc.
14$^{th}$ IUTAM Congress. Delft, Netherlands, 1976). W.T. Koiter, ed.,
North-Holland Publ. Comp., 1977.

Lin, C.C., Shu, F.H. 1964. \textit{On the spiral structure of disk galaxies}. ApJ \textbf{140}, 646-655.

Lin, C.C., Shu, F.H. 1966. \textit{On the spiral structure of disk galaxies. II. Outline of a theory of density waves}. Proc. Nat. Acad. Sci. \textbf{55}, 229-234.

Lin, C.C., Shu, F.H. 1967. \textit{Density waves in disk galaxies}. Noordwijk 1966, 313-317.

Lin, C.C., Shu, F.H. 1971. \textit{Density wave theory of spiral structure}. In: Brandeis University Summer Institute in
theoretical Physics, 1968. Astrophysics and General relativity, Vol. 2.
M.Chretien, S.Deser, J.Goldstein, eds. Gordon and Breach Science Publ., pp.
239-329.

Lin, C.C., Yuan, C., Shu, F.H. 1969. \textit{On the structure of disk galaxies. III. Comparison with observations}. ApJ \textbf{155}, 721-746 (LYS).

Lindblad, B. 1963. \textit{On the possibility of a quasi-stationary spiral structure in galaxies}. Stockholm Obs. Ann. \textbf{22}, No.5.

Lindblad, P.O. 1962. \textit{Gravitational resonance effects in the central layer of a galaxy.} Princeton 1961, 222-233.

Lynden-Bell, D. 1960. \textit{Stellar and Galactic Dynamics}. PhD Thesis, Univ. of Cambridge.

Lynden-Bell, D. 1962. \textit{Stellar dynamics. Potentials with isolating integrals}. MNRAS \textbf{124}, 95-123.

Lynden-Bell, D. 1964a. \textit{Letter to A. Toomre} (13 Jul).

Lynden-Bell, D. 1964b. \textit{Letter to A. Toomre} (Aug).

Lynden-Bell, D. 1964c. \textit{Letter to A. Toomre} (early Sep).

Lynden-Bell, D. 1964d. \textit{Letter to A. Toomre} (02 Dec).

Lynden-Bell, D. 1965a. \textit{Free precession for the galaxy}. MNRAS \textbf{129}, 299-307.

Lynden-Bell, D. 1965b. \textit{Letter to A. Toomre} (04 Dec).

Lynden-Bell, D. 1966.\textit{The role of magnetism in spiral structure}. Observatory \textbf{86}, No 951, 57-60.

Lynden-Bell, D. 1967. \textit{Statistical mechanics of violent relaxation in stellar systems}. MNRAS \textbf{136}, 101-121.

Lynden-Bell, D. 1969. \textit{Galactic nuclei as collapsed old quasars}. Nature \textbf{223}, 690-694.

Lynden-Bell, D. 1974. \textit{On spiral generating}. Proc. First European Astron. Meeting (Athens, Sept.
1972), \textbf{3}, Springer-Verlag, 114-119.

Lynden-Bell, D., Ostriker, J.P. 1967. \textit{On the stability of differentially rotating bodies}. MNRAS \textbf{136}, 293-310.

Lynden-Bell, D., Kalnajs, A.J. 1972. \textit{On the generating mechanism of spiral structure}. MNRAS \textbf{157}, 1-30.

Metzger, P.G. 1970. \textit{The distribution of HII regions}. Basel 1969, 107-121.

Miller, R.H., Prendergast, K.H. 1968. \textit{Stellar dynamics in a discrete phase space}. ApJ \textbf{151}, 699-709.

Miller, R.H., Prendergast, K.H., Quirk, W.J. 1970a. \textit{Numerical experiments in spiral structure}, Basel 1969, 365-367.

Miller, R.H., Prendergast, K.H., Quirk, W.J. 1970b. \textit{Numerical experiments on spiral structure}. ApJ \textbf{161},
903-916.

Oort, J.H. 1962. \textit{Spiral structure}. Princeton 1961, 234-244.

Oort, J.H. 1965. Discussion on the report\textit{ Some topics concerning the structure and evolution of galaxies}. In: The structure and evolution
of galaxies (Proc. 13$^{th}$ Conf. Physics, Univ. of Brussels, Sept. 1964.).
Interscience Publ., p. 23.

Oort, J.H. 1970. \textit{Survey of spiral structure problems}. Basel 1969, 1-5.

Pasha, I.I. 2002. \textit{Density-wave spiral theories in the 1960s}. I. Historical and astronomical researches, \textbf{27},
102-156 (Paper I, in Russian).

Pavlovskaya, E.D., Sharov, A.S. 1970. \textit{The Galactic structure and the appearance of the Milky Way}. Basel 1969, 222-224.

Pfleiderer, J. 1963. \textit{Gravitationseffekte bei der Begegnung zweier Galaxien}. Zeitschr. f. Astrophys. \textbf{58}, 12-22.

Pfleiderer, J., Siedentopf, H. 1961. \textit{Spiralstrukturen durch Gezeiteneffekte bei der Begegnung zweier Galaxien}. Zeitschr. f. Astrophys. \textbf{51},
201-205.

Prendergast, K.H. 1962. \textit{The motion of gas in barred spiral galaxies}. Princeton 1961, 217-221.

Prendergast, K.H. 1967. \textit{Theories of spiral structure}. Noordwijk 1966, 303-312.

Prendergast, K.H., Burbidge, G.R. 1960. \textit{The persistence of spiral structure.} ApJ \textbf{131}, 243-246.

Pikelner, S.B. 1965\textit{. Spiral arms and interacting galaxies}. Astron. Zh. \textbf{42}, 515-526.

Pikelner, S.B. 1968. \textit{Structure and dynamics of the interstellar medium}. Ann. Rev. Astron. Astrophys. \textbf{6}, 165-194.

Pikelner, S.B. 1970. \textit{Shock waves in spiral arms of Sc galaxies}. Astron. Zh. \textbf{47}, 752-759.

Roberts, W.W. 1968. \textit{Shock Formation and Star Formation in Galactic 
Spirals}. PhD thesis, MIT.

Roberts, W.W. 1969. \textit{Large-scale shock formation in spiral galaxies and its implications on star formation}. ApJ \textbf{158}, 123-143.

Roberts, W.W. 1970. \textit{Large-scale galactic shock phenomena and the implications on star formation}. Basel 1969, 415-422.

Roberts, W.W., Yuan, C. 1970. \textit{Application of the density-wave theory to the spiral structure of the Milky Way system. III. Magnetic field: large-scale hydromagnetic shock formation}. ApJ \textbf{161}, 877-902.

Roberts, W.W., Shu, F.H. 1972. \textit{The role of gaseous dissipation in density waves of finite amplitude}. Astrophys. Lett. \textbf{12}, 49-52.

Spitzer, L., Schwarzschild, M. 1951. \textit{The possible influence of interstellar clouds on stellar velocities}. ApJ \textbf{114}, 385-397.

Spitzer, L., Schwarzschild, M. 1953. \textit{The possible influence of interstellar clouds on stellar velocities}. II. ApJ \textbf{118}, 106-112.

Shu. F.H.-S. 1968. \textit{The Dynamics and Large-Scale Structure of Spiral Galaxies}. PhD Thesis, Dept. of Astron., Harvard Univ., Cambridge,
MA (January 1968).

Shu, F.H. 1969. \textit{Models of partially relaxed stellar disks}. ApJ \textbf{158}, 505-518.

Shu, F.H. 1970a. \textit{The propagation and absorption of spiral density waves}. Basel 1969, 323-325.

Shu, F.H. 1970b. \textit{On the density-wave theory of galactic spirals. I. Spiral structure as a normal mode of oscillation}. ApJ \textbf{160}, 89-97.

Shu, F.H. 1970c. \textit{On the density-wave theory of galactic spirals. II. The propagation of the density of wave action}. ApJ \textbf{160}, 99-112.

Shu. F.H. 2001. Private communication.

Shu, F.H., Stachnik, R.V., Yost, J.C. 1971. \textit{On the density-wave theory of galactic spirals. III. Comparisons with external galaxies}. ApJ \textbf{166}, 465-481.

Shu, F.H., Milione, V., Gebel, W., Yuan, C., Goldsmith, D.W., Roberts, W.W.
1972. \textit{Galactic shocks in an interstellar medium with two stable phases}. ApJ \textbf{173}, 557-592.

Simonson, S.C. III, 1970. \textit{Problems in galactic spiral structure: an account of a ``Spiral Workshop''}. Astron. Astrophys. \textbf{9}, 163-174.

Shane, W.W., Bieger-Smith, G.P. 1966. \textit{The Galactic rotation curve derived from observations of neutral hydrogen}. Bull. Astron. Inst. Netherl.
\textbf{18}, 263-292.

Stromgren, B. 1966a. \textit{Spectral classification through photoelectric narrow-band photometry}. Ann. Rev. Astron. Astrophys. \textbf{4}, 433-472.

Stromgren, B. 1966b. In \textit{Outline of talks presented at the Columbia November 4, 1966 Meeting, from notes taken by F. Shu}.

Stromgren, B. 1967. \textit{Places of formation of young and moderately young stars}. Noordwijk 1966, 323-329.

Tashpulatov, N. 1969. \textit{Tidal interaction of galaxies and eventual formation of junctions and ``tails''}. Astron. Zh. \textbf{46}, 1236-1246.

Tashpulatov, N. 1970. \textit{The possibility of the formation of bonds and tails as a result of tidal interaction of galaxies}. Astron. Zh. \textbf{47}, 277-291.

Thorne, R.M. 1968. \textit{An estimate of the enhancement of dynamical friction by stellar co-operative effects}. ApJ \textbf{151}, 671-677.

Toomre, A. 1964a. \textit{On the gravitational stability of a disk of stars}. ApJ \textbf{139}, 1217-1238 (T64).

Toomre, A. 1964b. \textit{Letter to D. Lynden-Bell} (07 Jul).

Toomre, A. 1964c. \textit{Letter to D. Lynden-Bell} (28 Aug).

Toomre, A. 1964d. \textit{Letter to D. Lynden-Bell} (14 Nov).

Toomre, A. 1965. \textit{On the gravitational instabilities of a galactic gas disk}. Submitted to ApJ (July 1965), unpublished.

Toomre, A. 1969. \textit{Group velocity of spiral waves in galactic disks}. ApJ \textbf{158}, 899-913 (T69).

Toomre, A. 1974. \textit{Gravitational interactions between galaxies}. In: The formation and the dynamics of galaxies. J.R.
Shakeshaft, ed., 347-365.

Toomre, A. 1970. \textit{Spiral waves caused by a passage of the LMC?} Basel 1969, 334-335.

Toomre, A. 1977. \textit{Theories of spiral structure}. Ann. Rev. Astron. Astrophys. \textbf{15}, 437-478.

Toomre, A. 1981. \textit{What amplifies the spirals?} In: The structure and Evolution of Galaxies. S.M. Fall and
D. Lynden-Bell, eds., Cambridge Univ. Press, 111-136.

Toomre, A., Toomre, J. 1972. \textit{Galactic bridges and tails}. ApJ \textbf{178}, 623-666.

Toomre, A., Toomre, J. 1973. \textit{Violent tides between galaxies}. Scientific Amer. \textbf{229}, 39-48.

Vaucouleurs, G. de, 1970. \textit{Statistics of spiral patterns and comparison of our Galaxy with other galaxies}. Basel 1969, 18-25.

Vorontsov-Velyaminov, B.A. 1962. \textit{Interaction of multiple systems}. In: Problems of Extra-Galactic Research
(IAU Symp. No.15). G.C. McVittie, ed., NY, Macmillan, 194-200.

Vorontsov-Velyaminov, B.A. 1964. \textit{Evidence of magnetic-like phenomena in the structure of galaxies.} Astron. Zh. \textbf{41}, 814-822.

Vorontsov-Velyaminov, B.A. 1970. \textit{Spiral structure of our Galaxy and of other galaxies}. Basel 1969, 15-17.

Weaver, H. 1970. \textit{Spiral structure of the Galaxy derived from the Hat Creek survey of neutral hydrogen}. Basel 1969, 126-139.

Woltjer, L. 1965. \textit{Dynamics of gas and magnetic fields; Spiral structure}. In: Galactic structure. A. Blaauw, M. Schmidt, eds.,
Univ. Chicago Press, 531-587.

Yuan, C. 1969a. \textit{Application of the density-wave theory to the spiral structure of the Milky-Way system. I. Systematic motion of neutral hydrogen}. ApJ \textbf{158}, 871-888.

Yuan, C. 1969b. A\textit{pplication of the density-wave theory to the spiral structure of the Milky-Way system. II. Migration of star.} ApJ \textbf{158}, 889-898.

Yuan, C. 1970. \textit{Theoretical 21-cm line profiles: comparison with observations}. Basel 1969, 391-396.

Zasov, A.V. 1967. \textit{On the possibility of many years' life of interstellar arms}. Astron. Zh. \textbf{44}, 975-980.

Zwicky, F. 1963. \textit{Intergalactic bridge.} ASP Leaflet No. 403, 17-24.

Zwicky, F. 1967. Adv. Astron. Astrophys. \textbf{5}, 267.

}
}
\end{document}